\documentclass[11pt,preprint]{aastex}
\usepackage{graphics}
\usepackage{amssymb}
\usepackage{amsmath}
\usepackage[dvips]{color}
\def\A{${\rm \AA}$}
\def\deg{$^{\mathrm o}$~}
\newcommand{\kms}{\ensuremath{\mathrm{km~s^{-1}}}}
\newcommand{\feii}{Fe\,{\footnotesize II}}
\newcommand{\mgii}{Mg\,{\footnotesize II}}
\newcommand{\heii}{He\,{\footnotesize II} }
\newcommand{\civ}{C\,{\footnotesize IV}}
\newcommand{\nv}{N\,{\footnotesize V}}
\newcommand{\niv}{N\,{\footnotesize IV}]}
\newcommand{\niii}{N\,{\footnotesize III}]}
\newcommand{\ovi}{O\,{\footnotesize VI}}
\newcommand{\oiii}{O\,{\footnotesize III}]}
\newcommand{\siiv}{Si\,{\footnotesize IV}}

\begin{document}

\title{On the scattering enhancement of \nv $\lambda$1240 emission line of quasi-stellar objects}
\author{Huiyuan Wang\altaffilmark{1,2}, Tinggui Wang\altaffilmark{1,2}, Weimin Yuan\altaffilmark{3,4}, Junxian Wang\altaffilmark{1,2}, Xiaobo Dong\altaffilmark{1,2} and
Hongyan Zhou\altaffilmark{1,2}}
 \altaffiltext{1}{Key Laboratory
for Research in Galaxies and Cosmology, University of Science and
Technology of China, Chinese Academy of Sciences, Hefei, Anhui
230026, China}
\altaffiltext{2}{Center for Astrophysics,
University of Science and Technology of China, Hefei, Anhui
230026, China}
\altaffiltext{3}{National Astronomical
Observatories/Yunnan Observatory, Chinese Academy of Sciences,
P.O. Box 110, Kunming, Yunnan 650011, China} \altaffiltext{4}{Key
Laboratory for the Structure and Evolution of Celestial Objects,
Chinese Academy of Sciences, P.O. Box 110, Kunming, Yunnan 650011,
China}
\email{whywang@mail.ustc.edu.cn}

\begin{abstract}
The \nv\ emission line of active galactic nuclei shows peculiar
behavior in the line--continuum correlation, which may be
indicative of an extra line component in addition to that from the
normal broad emission line region. In this paper, we investigate
possible contribution to the \nv\ emission via resonant scattering
of both continuum and Ly$\alpha$ in a broad absorption line (BAL)
outflow, by performing the Sobolev Monte Carlo simulations. The
contribution is dependent on the covering factor, optical depth
and velocity profile of the outflow, as well as the equivalent
width (EW) of Ly$\alpha$. Adopting model parameters constrained by
observations, we find that the measured \nv\ EW in the spectra of
non-BAL quasi-stellar objects (QSOs) could have been enhanced by a
factor of 1.82$-$2.73 on average, while there is only moderate
absorption of Ly$\alpha$ \emph{along the BAL outflow direction}.
Our model can produce a relatively narrow scattering line profile.
About 80\% of the total scattered flux falls within the central
$\pm4500$\kms. We find that the resonant scattering can produce a
prominent polarized emission line around \nv. Both the broad
excess emission and the unusually large polarized flux observed
around \nv\ in BAL QSOs are considered as strong evidence for the
scattering enhancement. Future spectropolarimetric observations
and spectroscopic monitoring of luminous QSOs may offer crucial
tests for this interpretation, and provide useful information on
the physical and geometrical properties of QSO outflows. We argue
that the scattering offers a promising and robust process for
producing the peculiar behavior of \nv\ emission compared to the
other processes proposed previously.
\end{abstract}


\section{Introduction}

Much of our knowledge about active galactic nuclei (AGNs) and the
more luminous quasi-stellar objects (QSOs) is derived from their
emission lines, especially the broad emission lines (BELs), in the
optical and ultraviolet band. Photoionization models,
incorporating the treatment of ionization, thermal equilibrium,
and the radiation transfer process, can reproduce successfully the
overall observed emission line spectra with only a few exceptions,
such as the \feii\ and \nv\ emission lines (e.g., Rees et al.
1989; Baldwin et al. 1995; Ferland et al. 1998). Studies of the
emission lines improved our understanding of the physical state
and chemical compositions of the BEL gas, and further the central
engine of AGNs. In particular, the abundant ultraviolet lines,
such as the prominent metal lines \ovi$\lambda1035$,
\nv$\lambda1240$, \siiv$\lambda1397$ and \civ$\lambda1549$, are
valuable probes of high-redshift QSOs in the early Universe owing
to the unique information provided by these lines.

These metal lines are commonly thought to be emitted in dense
photoionized clouds via collisional excitation (e.g., Peterson
1997; Osterbrock \& Ferland 2006; cf. Murray \& Chiang 1997).
Among them, the \nv\ BEL is very special. Its equivalent width
(EW) is nearly independent of the continuum luminosity (e.g.,
Osmer et al. 1994; Dietrich et al. 2002), whereas those of the
other high ionization lines (e.g., \civ\ and \ovi) show
significant anti-correlations with the continuum luminosity, known
as the Baldwin effect (Baldwin 1977; Laor et al. 1995; Zheng et
al. 1995; Wang et al. 1998; Dietrich et al. 2002; Dong et al.
2009). Compared to other lines, \nv\ is particularly strong in the
spectra of high luminosity QSOs. This leads to the suggestion that
luminous QSOs have higher metallicity gas than less luminous QSOs,
assuming that \nv\ is collisionally excited (e.g., Hamann \&
Ferland 1993;1999; Shemmer \& Netzer 2002).

However, it has also been suggested that some of the observed \nv\
emission could result from resonant scattering of photons of both
the continuum and the Ly$\alpha$ emission line in the broad
absorption line region (BALR; Surdej \& Hutsem$\grave{e}$kers
1987; Turnshek et al. 1988; Weymann et al. 1991; Hamann et al.
1993, hereafter HKM; Hamann \& Korista 1996, hereafter HK96;
Krolik \& Voit 1998; Wang et al. 2007, hereafter WWW). Broad
absorption lines (BALs), which are produced in partially ionized
outflows with typical velocities of 10000--20000\kms, are observed
in the ultraviolet spectra of about 10\%--20\% luminous QSOs
(Weymann et al. 1991; Hewett \& Foltz 2003; Knigge et al. 2008).
At such velocities N$^{+4}$ ions in the BALR, albeit its small
covering factor (Weymann et al. 1991; Reichard et al. 2003), can
scatter photons of Ly$\alpha$ and the continuum into the
wavelengths around \nv, resulting in the observed enhancement of
the \nv\ emission. Indeed, absorption of Ly$\alpha$ photons by
N$^{+4}$ ions has been observed in many BAL QSOs (e.g., Turnshek
et al. 1988). Since such a BAL outflow is expected to exist in
essentially every QSO, the scattering enhancement of the \nv\ BEL
should be a commonplace in QSOs and not exclusive of BAL QSOs.

However, the significance or the strength of the scattering
emission is still under debate. On one hand, Surdej \&
Hutsem$\grave{e}$kers (1987) estimated that the \nv\ flux can be
enhanced by a factor of several (see also Krolik \& Voit 1998), by
simply assuming that the majority of Ly$\alpha$ photons along the
BAL outflow direction are scattered and assuming a large covering
factor of the outflow. On the other hand, using an optical depth
profile derived from the mean spectrum of a sample of BAL QSOs,
Hamann and Korista (1996) found that only 30\% of the incident
Ly$\alpha$ photons are scattered by the BAL gas. Furthermore,
scattering emission from a high-velocity outflow is generally
broader than typical BEL, and thus only a small fraction of the
scattered photons can contribute to the measured \nv\ EW. They
thus reached a different conclusion that the scattering accounts
for no more than 18\% of the measured \nv\ emission only.

In this paper, we re-examine this issue by making use of sets of
improved parameters on the geometry and optical depth of BAL
outflows as constrained by recent observations. We show that the
scattering contribution to measured \nv, which is strongly model
parameter dependent, can be considerable for the reasonable values
of the parameters. Our model is described in detail in SEC.
\ref{sec_model}. The primary results are presented in SEC.
\ref{sec_enh}, and are discussed in SEC. \ref{sec_ori} and
\ref{sec_pro} by addressing issues such as what causes the
differences between the previous results and ours. Observational
evidence in support of the scattered emission is presented
 in SEC. \ref{sec_evi}. We try to explain the  peculiar behavior of \nv\ based on our
model in SEC. \ref{sec_nv}. We investigate the effect of differing
the geometric model on the scattering enhancement in SEC.
\ref{sec_fun}. The conclusion is summarized at the end.

\section{Model Description}\label{sec_model}

We use the numerical algorithm developed by WWW to simulate the
resonant scatter process in an outflow with a large velocity
gradient and terminal velocity. The Monte Carlo technique and the
Sobolev approximation are adopted for the treatment of line
transfer. These two techniques have been widely used to deal with
radiation transfer in various media (e.g., Lee et al. 1994;
Goosmann \& Gaskell 2007). Different from the escape probability
method (e.g., HKM), which can only approximate resonant scatter
for a singlet transition, our code can handle multiple
transitions. Moreover, our code can calculate the polarization of
scattered light. We refer readers to WWW for more details (see
also Lee et al. 1994; Lee 1994; Lee \& Blandford 1997). Here we
only describe briefly our assumptions and model parameters used in
this paper.

\begin{figure}
\plotone{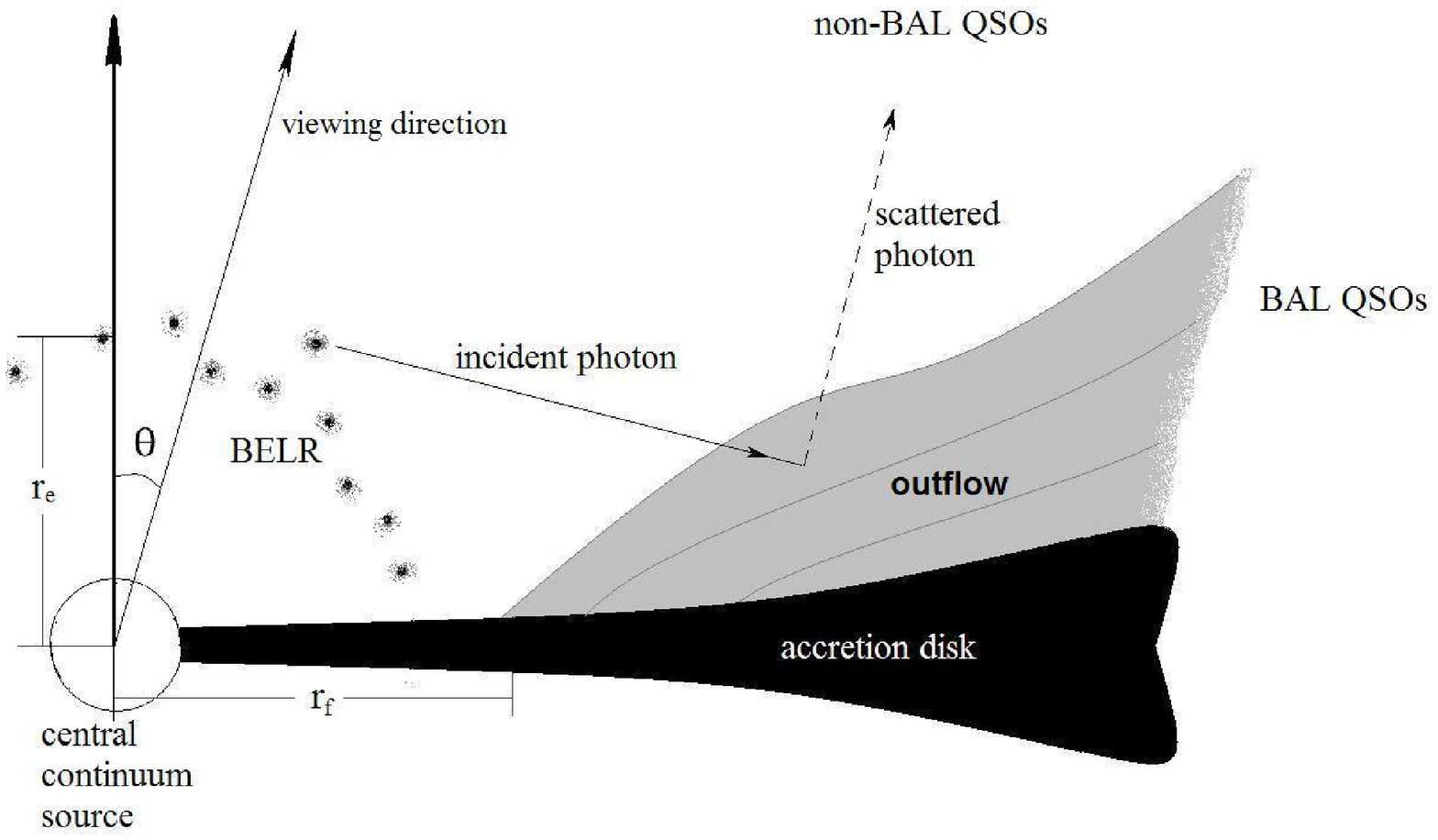}
 \caption{\label{fig_dw} Cartoon sketch of the cross section of the adopted geometry of an accretion disk, a BELR and
 outflow around a central engine. The black region represents the accretion disk,
 and the the open circle represents the central continuum source.
 Outflow is shown as the shaded area with a few streamlines.
 The scattered dots  represent the BELs clouds.
 The thick line with an arrow denotes the rotation axis of the disk.}
\end{figure}

We consider an equatorial and axisymmetric outflow that is
supposed to launch initially from an accretion disk and then
radially accelerated by the radiation pressure (e.g., Murray et
al. 1995; hereafter M95; de Kool \& Begelman 1995; Proga et al.
2000). Complications to this simple model and their implication
are discussed in SEC. \ref{sec_fun}. When the line of sight to the
continuum source intersects the outflow, BALs are produced in the
QSO spectrum; alternatively, the QSO appears as a non-BAL QSO
(Fig. \ref{fig_dw}). Thus the incidence of the BALs in QSOs
spectra is a measure of the average covering factor (CA) of the
BALR. Although the exact value is still uncertain owing to various
selection effects in QSO samples (e.g., Krolik \& Voit 1998;
Hewett \& Foltz 2003) and different criteria for BAL QSOs (e.g.,
Knigge et al. 2008), more recent studies suggest a BAL fraction of
around 20\%, almost a factor of 2 larger than the previous value
of 10\% (Hewett \& Foltz 2003; Reichard et al. 2003; Dai et al.
2008; Knigge et al. 2008). There is also evidence for a wide range
of the BALR covering factor among QSOs, whose value may be related
to the fundamental parameters of QSOs, such as luminosity, black
hole mass and accretion rate (Morris 1988; Ganguly et al. 2007).
To be representative, we consider two values of the covering
factor, CA$=0.1$ and 0.2, in this paper.

We adopt M95's formalism to describe the radial velocity
distribution: $v_r=v_t(1-r_f/r)^\beta$, where $v_t$ is the
terminal velocity and $r_f$ is the launch radius of the outflow.
$v_t$ and the initial velocity $v_0$ of the absorption trough are
fixed at 15000\kms~and 1000\kms, respectively\footnote{Our adopted
initial velocity of BAL is similar to what HK96 adopted,
especially their extreme case}. Given an average optical depth
over the flow velocity, $\tau_0$, one can derive the radial
optical depth as a function of the radial velocity $v_r$,
\begin{eqnarray}
\tau_r (v_r) = \frac{\Delta
v}{v_r}\frac{\alpha\tau_0}{(v_r/v_0)^{\alpha}-(v_r/v_\infty)^{\alpha}}
\end{eqnarray}
where $\alpha=1-1/\beta$, $\Delta v=v_t-v_0$(see WWW for the exact
definition of $\tau_0$ and the derivation of this formula). The
optical depth profile is determined by two parameters, $\beta$ and
$\tau_0$. In principle, $\tau_r(v_r)$ can be obtained directly
from observed absorption line profile of the correspondent ions,
and can thus be used to test our assumption on the velocity field
and to determine $\tau_0$ and $\beta$, as done by HK96. In
practice, due to partial covering and/or filling of the scattered
light, the residual flux in the absorption trough can not be
directly used to determine the optical depth(Arav et al. 1999a).
Following Murray \& Chiang (1997), we consider several $\beta$
values between 0.5 and 2.0. We use $\tau_0=$5 as a representative
value. Although this value is much larger than the apparent
optical depth directly derived from the absorption trough (e.g.,
HK96), it is more consistent with recent detailed analysis (Hamann
1998; Arav et al. 1999a; Wang et al. 1999; Lu et al. 2008). For
comparisons, the results obtained with other values of $\tau_0$
are also presented.

The geometry and size of the continuum source and broad emission
line region (BELR) are also relevant to our calculation. The
central continuum source can be considered to be point-like with
respect to the BALR, based on previous studies (e.g., HKM; HK96;
WWW). Although the relative size of the BELR and the outflow is
not well determined from observations, there are several important
constraints. First, the Ly$\alpha$ emission line is partially
absorbed by the \nv\ BAL, suggesting that the BELR must lie within
or be cospatial with the BALR (e.g., Turnshek et al. 1988).
Second, BELs are much less polarized than the continuum in BAL
QSOs. This implies that the BELR should not be much smaller than
the electron scattering region (Goodrich \& Miller 1995; Schmidt
\& Hines 1999; Ogle 1997). Finally, the very large ($>$10\%)
polarization detected in the CIV BAL trough in some QSOs hints at
that the electron scattering region has a size comparable to that
of the BALR. Due to the close doublet nature, the resonantly
scattered light in an accelerated outflow is only moderately
polarized. Thus, the electron scatted light that does not pass the
outflow is likely to be responsible for the high polarization
degree (Goodrich \& Miller 1995; Ogle et al. 1999; Lamy \&
Hutsem$\grave{e}$kers 2004; WWW). To meet all these constraints,
the size of the BELR must be the same order of magnitude as that
of the BALR. A recent study based on polarized features across the
H$\alpha$ emission line in the BAL QSO PG 1700+158 provides
convincing evidence that the outflow is launched from a region
roughly cospatial with the BELR (Young et al. 2007). It is
therefore reasonable to assume that the emission line clouds are
distributed within a thin spherical shell with a radius comparable
to the inner radius of the outflow (see Fig. \ref{fig_dw}). In
this paper, the radius of the \nv\ emission region is fixed to
$0.2r_f$\footnote{Our additional tests show that the results are
insensitive to the size of the \nv\ emission region. We thus fix
it for simplicity.}, and several values of the radius of the
Ly$\alpha$ emission region ($r_e$) are considered.

Both the  strength and profiles of the emission lines are
important input parameters for our simulations. Due to severe
Ly$\alpha$ forest absorption and heavy blend with \nv, it is
difficult to obtain precisely the EW of Ly$\alpha$. Vanden Berk et
al. (2001) measured a Ly$\alpha$ EW of 92.9\A\ from a composite
spectrum of Sloan Digital Sky Survey (SDSS) QSOs, while HK96
obtained a smaller value of 66\A. In the simulations below, we
adopt the EWs of the intrinsic emission lines\footnote{Here, we
refer to emission lines from the normal BELR as intrinsic emission
lines.} as EW$_{\rm sim}$ = 80\A~for Ly$\alpha$ and 10\A~for \nv,
and their profiles as a Gaussian with FWHM$=$4000\kms. The adopted
value for \nv\ is less than what HK96 used, 18\A, since the \nv\
emission is enhanced by resonant scattering according to our
calculation in this paper. Moreover, the scattering enhancement is
insensitive to the intrinsic \nv\ EW value (see below). Although
\nv\ consists of a doublet with a separation of 4\A, we
approximate the \nv\ profile with a single Gaussian. We expect
that this has little effect on the results obtained in this paper.
The continuum spectrum is parameterized by a power law with an
index of zero. Both the continuum and line emission are assumed to
be isotropic.

\begin{table}
\begin{center}
\caption{The Model Parameters and Average Absorption EWs}
\label{tab_ew}
\begin{tabular}{lccccc}
  \hline\hline
  ~~~~~~~~& Model Parameters& ~~~~~~~~~~~~~ & $\overline{{\rm EW}}_{\rm ab}$(\A) &~~~~~~ \\
\end{tabular}
\begin{tabular}{lccccccccc}
  \cline{2-5}\cline{7-9}
  No. & CA & $\beta$ & $\tau_0$ & $r_e/r_f$ & & Continuum & Ly$\alpha$ & \nv \\
  \hline
  1 & 0.1 & 2.0 & 5 & 0.5 & & 50.5  & 86.2  & 4.6\\
  2 & 0.2 & 2.0 & 5 & 0.5 & & 50.3  & 82.8  & 4.4\\
  3 & 0.2 & 2.0 & 5 & 0.0 & & 50.3  & 75.8  & 4.4\\
  4 & 0.2 & 2.0 & 5 & 1.0 & & 50.3  & 89.4  & 4.4\\
  5 & 0.2 & 1.5 & 5 & 0.5 & & 52.9  & 84.7  & 4.4\\
  6 & 0.2 & 1.0 & 5 & 0.5 & & 56.4  & 87.2  & 4.3\\
  7 & 0.2 & 0.5 & 5 & 0.5 & & 59.6  & 88.9  & 3.9\\
  8 & 0.2 & 2.0 &10 & 0.5 & & 56.5  & 88.4  & 4.4\\
  9 & 0.2 & 2.0 & 3 & 0.5 & & 43.6  & 73.4  & 4.4\\
  10 & 0.2 & 2.0 & 1.5 & 0.5 & & 33.4  & 54.8  & 4.2\\
  11 & 0.2 & 2.0 & 0.5 & 0.5 & & 18.2  & 26.3  & 3.1\\
  \hline
\end{tabular}
\tablecomments{There are $\overline{{\rm EW}}_{\rm
ab}=\overline{{\rm EW}}_{\rm sc}/{\rm 0.5CA}$. The additional
factor of 0.5 is introduced because half of scattered photons are
absorbed by the accretion disk in our model.}
\end{center}
\end{table}

Below we perform Monte Carlo simulations to model observed fluxes
around the Ly$\alpha$ and \nv\ emission lines as functions of
wavelength and orientation. The input parameters are listed in
Tab.\,\ref{tab_ew}. Then we investigate the contribution of
scattered light to the measured \nv\ emission. We note that, in
our model, photons encountering the equatorial plane (accretion
disk) would be absorbed, i.e. the accretion disk is assumed to be
optically thick without reflection. This treatment is different
from some of the others and gives a conservative estimate of
scattering contribution.

\section{Enhancement of \nv\ Emission Line}\label{sec_enh}

\begin{figure}
\plotone{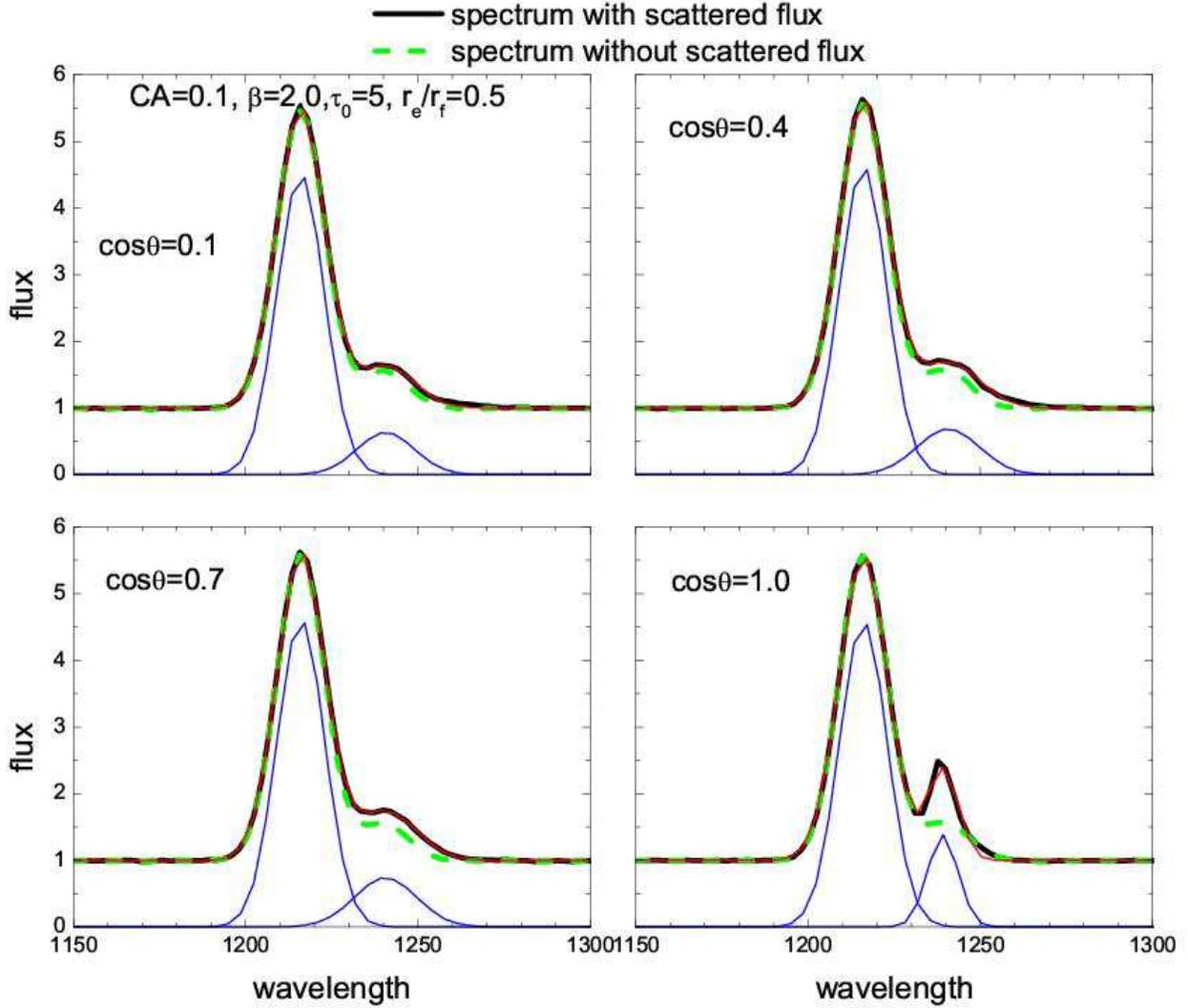}
 \caption{\label{fig_f10} Emerging spectra, with (solid, black lines) and without the scattered flux
 (dashed, green lines), along different directions denoted in each panel. The
best-fit Gaussian profiles are also shown (blue thin lines). The
red thin lines are the composite profiles of the two Gaussian
profiles and continuum. The model parameters are shown in the
top-left panel.}
\end{figure}

\begin{figure}
\plotone{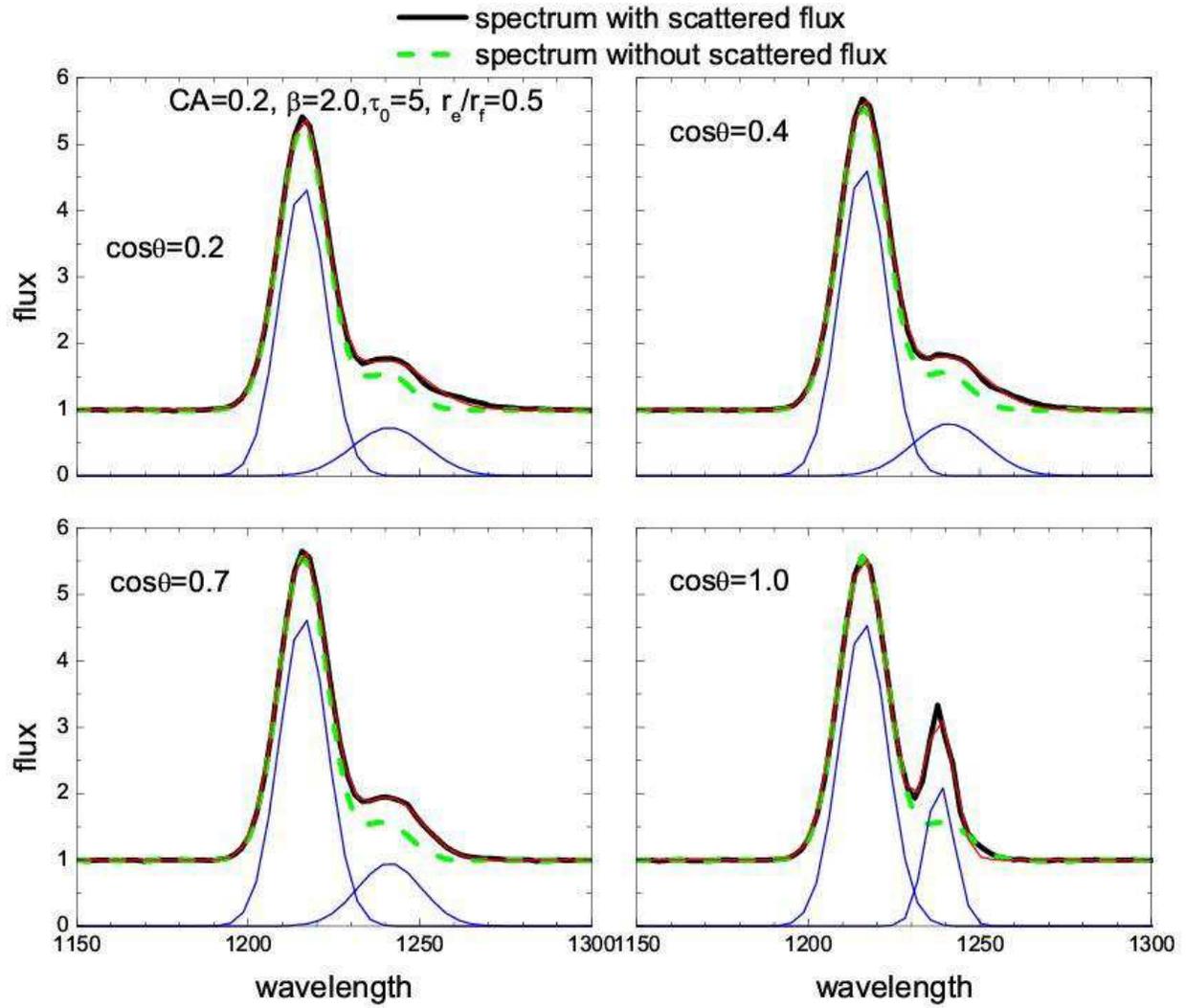}
 \caption{\label{fig_f20} Similar to Fig. 2 but a with covering factor of 0.2.}
\end{figure}

To give a visual impression of the scattering contribution to the
\nv\ BEL, we show in Fig. \ref{fig_f10} and \ref{fig_f20} the
emerging spectra with and without the scattered light, viewed
along several BAL-free directions (the line of sight to the
continuum source would not be intersected by the outflow)
specified by $\cos\theta$. Here, $\theta$ is the angle between the
viewing direction and the rotational axis of the equatorial
outflow (Fig. \ref{fig_dw}). The model parameters in Fig.
\ref{fig_f10} are CA=0.1, $\beta=2.0$, $\tau_0=5$ and $r_e=0.5r_f$
(Model 1 in Tab. \ref{tab_ew}), while those in Fig. \ref{fig_f20}
are similar to Model 1 except a larger CA of 0.2 (Model 2). In
both the models, the profile around the \nv\ line becomes narrower
as $\cos\theta$ increases. This is natural because the projected
velocity of the outflow along the polar direction is smaller than
that in the equatorial direction. However the profile of the
scattered photons seems to be more center-peaked in comparison of
those  in WWW (e.g., figure 24). The difference is a consequence
of two processes. First, in this paper, we take into account the
scattering of Ly$\alpha$, which offsets by only $\sim$5900\kms\ to
the blue side of \nv. Second, the separation of the \nv\
multiplets is wider than that of \civ\ considered by WWW, and the
radiation transfer effect tends to make the line narrower (refer
to SEC. \ref{sec_pro} for more details). We attempt to fit the
output line profile with two Gaussians, one for Ly$\alpha$ and
another for \nv. The fitting results are shown in the same figures
for comparison. With a somewhat surprise, they give rather good
fits to the Ly$\alpha$ and \nv\ profiles, indicating that the
intrinsic plus the scattered emission can also yield a profile
similar to the normal broad lines.

\begin{figure}
\plotone{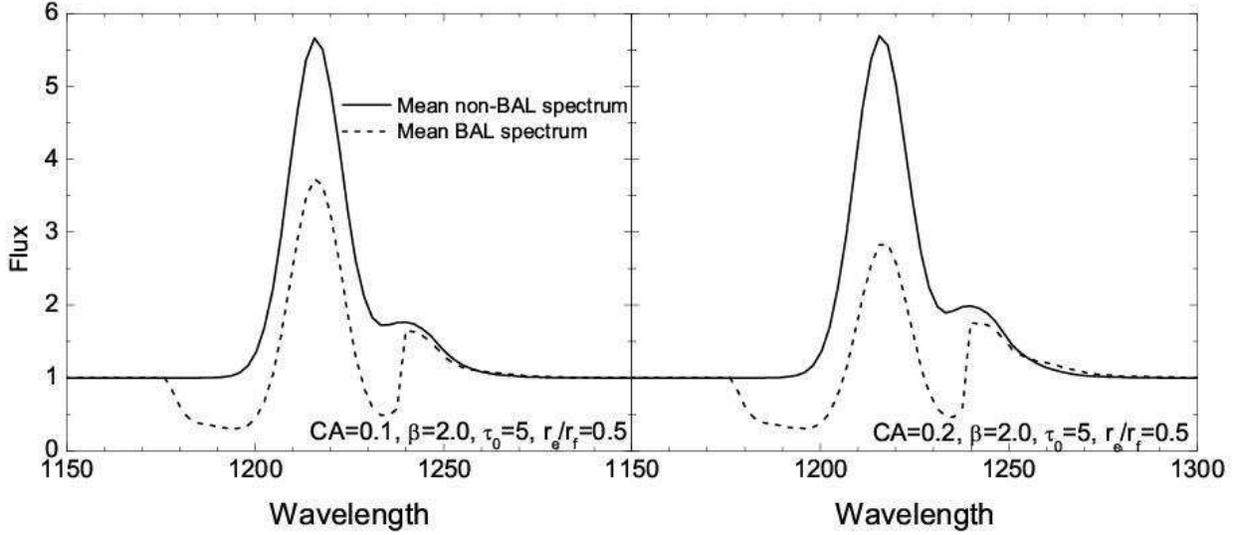} \caption{\label{fig_bal} The mean spectra of
BAL QSOs and non-BAL QSOs. The model parameters are shown in the
panels.}
\end{figure}

We show the angle-averaged spectra of BAL and non-BAL QSOs for the
two models in Fig. \ref{fig_bal}. As expected, only a fraction of
Ly$\alpha$ photons along the BAL direction are scattered by
N$^{+4}$ ions. The absorption EW to Ly$\alpha$ BEL is about 13.7
(32.0)\A\ for Model 1(2). This means that, on average, 17\%(40\%)
of Ly$\alpha$ photons initially along the BAL direction are
absorbed and re-emitted. This fraction is similar to that in HK96,
which adopted a small apparent optical depth instead of a large
optical depth plus partial coverage. Note that the difference in
the Ly$\alpha$ peak height between BAL and non-BAL QSOs is mainly
due to continuum absorption.

\begin{figure}
\plotone{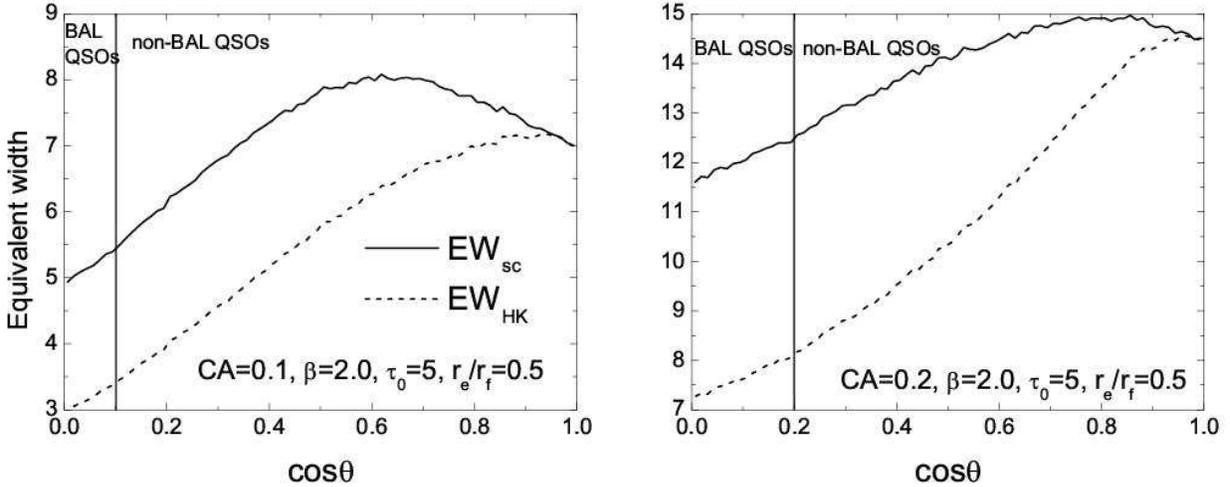}
 \caption{\label{fig_ew} The EW of the total scattering emission, EW$_{\rm sc}$, and the
 scattering emission in a certain velocity range (see text for details), EW$_{\rm HK}$,
 as a function of $\cos\theta$. The vertical lines denote
 the boundary between BAL and non-BAL QSOs. The model
parameters are shown in the panels.}
\end{figure}

Fig. \ref{fig_ew} shows the EW of the total scattering emission,
referred to as EW$_{\rm sc}$, as a function of $\cos\theta$. The
dependence of EW$_{\rm sc}$ on $\cos\theta$ is not monotonic, and
its turning point varies with CA. This behavior can be ascribed to
the large velocity splitting between the \nv\ doublet lines. We
will address this issue in details in SEC. \ref{sec_pro}. In a
direction free of BAL, EW$_{\rm sc}$ ranges from 5.5 (12.6) to 8.1
(15.0)\A, with a mean of 7.3 (14.2)\A, for the model with CA=0.1
(0.2).  As comparisons, the mean values in the BAL direction are
5.2 and 12.1\A\, respectively, for the two models. However, these
values should not be compared directly with the \nv\ EWs reported
in the literature. Because \nv\ is heavily blended with
Ly$\alpha$, some de-blending techniques are used to determine the
\nv\ EW. For instance, HK96 measured the \nv\ EW by integrating
the fluxes between -5000 and +4000\kms of the \nv\ line center in
excess of a symmetric fit to the Ly$\alpha$ profile. Other
de-blending methods, such as using the profiles of \civ\ as a
template to fit Ly$\alpha$+\nv\ or directly applying
multi-Gaussian component fitting, are also often used (e.g.,
Dietrich et al. 2003; Shemmer \& Netzer 2002). In order to compare
with the HK96 results, we measure the EWs of the scattering
emission between -5000 to +4000\kms centered at NV (referred to as
EW$_{\rm HK}$), as shown in Fig. \ref{fig_ew} (dashed lines). It
can be seen that EW$_{\rm HK}$ increases with $\cos\theta$ from
3.4(8.2) to 7.2(14.6)\A, with a mean value of 5.8 (11.4)\A. Even
for such a method, the contribution of the scattered light to the
\nv\ emission is still prominent. HK96 derived a \nv\ EW of 18\A\
from a composite spectrum of non-BAL QSOs. Based on our
calculations, the observed \nv\ emission is enhanced by an average
factor of 18/(18-5.8)$\simeq $1.48 or 18/(18-11.4)$\simeq$2.73,
depending on which CA value is chosen.

Even if the effect of absorption by the accretion disk is
considered and a slightly smaller covering factor is adopted, say,
CA$=$0.1 (HK96 used a CA=0.12), a larger enhancement than the HK96
value ($\leq18\A\times0.18\simeq3.2$\A) is still yielded.  The
discrepancy might result from the fact that we use a larger input
EW of Ly$\alpha$ and do not consider any partial covering of the
continuum emission, which is thought to be common (Arav et al.
1999a). To check the impact of these effects on the results, we
describe the EW of a scattering line in terms of the absorption
EW. Conservation of photons suggests that the average EW of the
scattered light ($\overline{{\rm EW}}_{\rm sc}$) equals the
average absorption EW ($\overline{{\rm EW}}_{\rm ab}$) times the
covering factor in the case of no absorption (see HK96). In our
model, an additional factor of 0.5 is introduced and the relation
reads as $\overline{{\rm EW}}_{\rm sc}$=0.5CA$\times\overline{{\rm
EW}}_{\rm ab}$, because half of the scattered photons are absorbed
by the accretion disk. Note that $\overline{{\rm EW}}_{\rm sc}$
here is the average over all the $4\pi$ solid angle rather than
over BAL-free directions  only. We separate the incident emission
into three components: continuum, Ly$\alpha$ and \nv~photons. Tab.
\ref{tab_ew} lists the $\overline{{\rm EW}}_{\rm ab}$ of the three
components.

The derived absorption EWs are insensitive to CA, at least for
CA$\leq0.2$. We can thus predict $\overline{{\rm EW}}_{\rm sc}$
from various EWs of the intrinsic emission lines, EW$_{\rm in}$,
and the covering factors via simple relations,
\begin{eqnarray}
\nonumber
 {\rm \overline{EW}_{sc}} &=& {\rm \frac{1}{2}CA[\overline{EW}_{ab}(
cont)}f_p+{\rm \frac{\overline{EW}_{ab}(Ly\alpha)}{EW_{sim}(
Ly\alpha)}EW_{in}(Ly\alpha)}\\
& & + {\rm \frac{\overline{EW}_{ab}(\nv)}{
EW_{sim}(\nv)}EW_{in}(\nv)}]\label{eq_calew}
\end{eqnarray}
where EW$_{\rm sim}$ denotes the EW of each emission line used in
our simulations: 80\A\ for Ly$\alpha$ and 10\A\ for \nv; $f_p$ is
the covering factor of the continuum source, which is used to
account for the partial coverage of the continuum. For a model
with $\beta=2.0$, $\tau_0=5$ and $r_e/r_f=0.5$, we have
\begin{equation}
{\rm\overline{EW}_{sc}=\frac{1}{2}CA}[50.4f_p+{\rm
1.06EW_{in}(Ly\alpha)+0.45EW_{in}(\nv)}]
\end{equation}
To compare with HK96 we adopt their parameters, CA=0.12 and ${\rm
EW_{in}(Ly\alpha)}$=66\A. HK96 measured a \nv\ EW of 18\A, which
is indeed the sum of the EWs of the intrinsic and scattering
emission. We thus adopt ${\rm EW_{in}(\nv)}=
18\A-\overline{EW}_{sc}$. HK96 determined the average continuum
absorption EW as 27\A\ in their case A from the absorption line
profiles; this corresponds to $f_p=0.54$. The resulting mean EW of
the total scattering emission is about 6.1\A, still slightly
larger than what HK96 obtained ($\leq5.6$\A).

\begin{figure}
\plotone{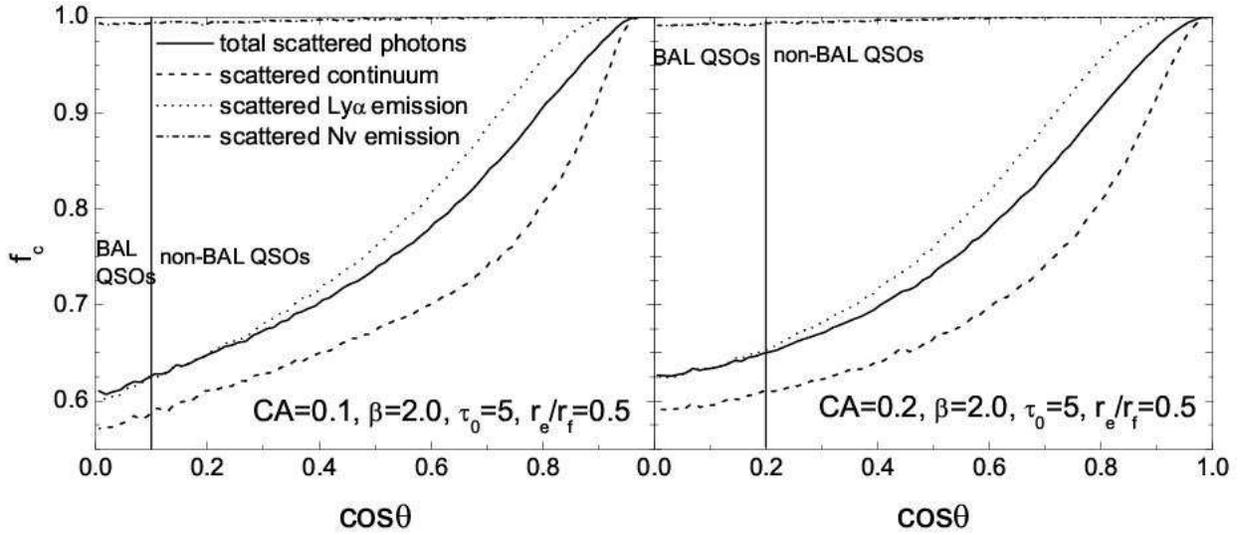}
 \caption{\label{fig_fc} $f_c$ parameter, which is defined as the ratio of EW$_{\rm
 HK}$ to EW$_{\rm sc}$, as a function of $\cos\theta$ for two models as indicated in the panels.
Different lines represent the results for total scattered photons,
scattered continuum, scattered Ly$\alpha$ emission line and
scattered \nv\ emission line, respectively. The vertical lines
denote the boundary between BAL and non-BAL QSOs.}
\end{figure}

To translate $\overline{{\rm EW}}_{\rm sc}$ into $\overline{{\rm
EW}}_{\rm HK}$ we define a parameter $f_c={\rm EW}_{\rm HK}/{\rm
EW}_{\rm sc}$. We show  in Fig. \ref{fig_fc} the $f_c$ parameter
for the total scattered photons (solid lines) as a function of
$\cos\theta$ for models 1 and 2. For comparisons, the results for
the scattered continuum, the scattered Ly$\alpha$ and \nv\
emission lines are also shown in different line styles as
indicated in the figure (Fig. \ref{fig_fc}). In a pure radial
outflow, the scattered emission from a higher velocity region is
broader than that from a lower velocity region. Continuum photons
can be scattered by N$^{+4}$ ions in the highest velocity region,
while most of the intrinsic Ly$\alpha$ and \nv\ photons can only
be scattered by outflow gas with radial velocity around 5900\kms
and $\lesssim 4000$\kms respectively. So the $f_c$ parameter is
the highest for intrinsic \nv\ emission followed by (in order)
Ly$\alpha$ and continuum emission. The $f_c$ parameter for the
total scattered photons increases with $\cos\theta$ because the
scattering emission line becomes broader at a larger polar angle.
On average, about 80\% of the total scattering line in BAL-free
directions contributes to EW$_{\rm HK}$. Adopting HK96's intrinsic
EWs and covering factor, the increased \nv~EW is about
$6.1\times0.8\approx4.9$\A, significantly larger than HK96's
result of $\leq3.2$\A. The \nv\ emission is on average enhanced by
a factor of 1.37. If adopting more representative CA of 0.2, the
enhancement factor increases to 1.82. We note again that about
half of the scattered photons are absorbed in our model. If the
absorption by the accretion disk is not important, the enhancement
would increase significantly.

Using Eq. \ref{eq_calew} we can also estimate the scattered flux
around the \civ\ emission line by assuming the same optical depth
profile for \civ\ as for \nv. Adopting a representative EW of
30\A\ for the intrinsic \civ\ emission line (Vanden Berk et al.
2001; HK96), the resulting EW$_{\rm HK}$ is 1.6(3.3)\A\ for
CA=0.1(0.2). The scattering emission can account for only
5.3\%--10.9\% of the observed \civ\ emission, and is thus not
important.

Given wide distributions of both the EW of the intrinsic emission
lines and the covering factor of the outflow, one can imagine a
very broad distribution of the scattered emission around \nv\ (Eq.
\ref{eq_calew}). In other words, while the \nv\ emission can be
enhanced by the scattered emission up to several times in some
QSOs, the enhancement may be negligible in others. According to
our calculations, the scattered flux is  dependent not only on the
EWs and covering factor, but also on the inclination. EW$_{\rm
HK}$ can differ by a factor of 2 at two extreme inclinations (Fig.
\ref{fig_ew}).

In comparison with HK96, a question arises as to why our model can
produce more contribution to \nv\ from scattered emission. We try
to answer this question from two aspects as follows. We discuss
where the seed photons of the scattering emission come from in
SEC. \ref{sec_ori}, and  the processes that shape the profiles of
the scattered light in SEC. \ref{sec_pro}.

\section{Origin of Scattered Photons}\label{sec_ori}

There seems to be a contradiction between the strong scattering
emission and the weak absorption of the Ly$\alpha$ emission shown
in the mean spectrum of BAL QSOs. From Tab. \ref{tab_ew} and Eq.
\ref{eq_calew}, one can find that the scattered flux is dominated
by scattered Ly$\alpha$ photons, while the listed $\overline{{\rm
EW}}_{\rm ab}$(Ly$\alpha$) is much larger than that derived from
the BAL profile (SEC. \ref{sec_enh}), and even larger than the
input emission EW 80\A\ for some models. All these can be readily
understood once the size of the Ly$\alpha$ emission region is
taken into consideration. In the case of a large relative size of
the BELR, the solid angle subtended by the outflow as viewed from
a BEL cloud is larger compared to that for a point-like BELR, so
N$^{+4}$ ions can scatter line photons which are traveling along
BAL-free directions. In the current model, Ly$\alpha$ photons
heading toward the accretion disk may be scattered by the outflow,
and eventually escape the BALR along the line of sight (see Fig.
\ref{fig_dw} for illustration). These Ly$\alpha$ photons provide
seed photons for resonant scattering in addition to the line
photons initially traveling along the BAL direction, and hence
strongly enhance the scattered flux around \nv.

\begin{figure}
\plotone{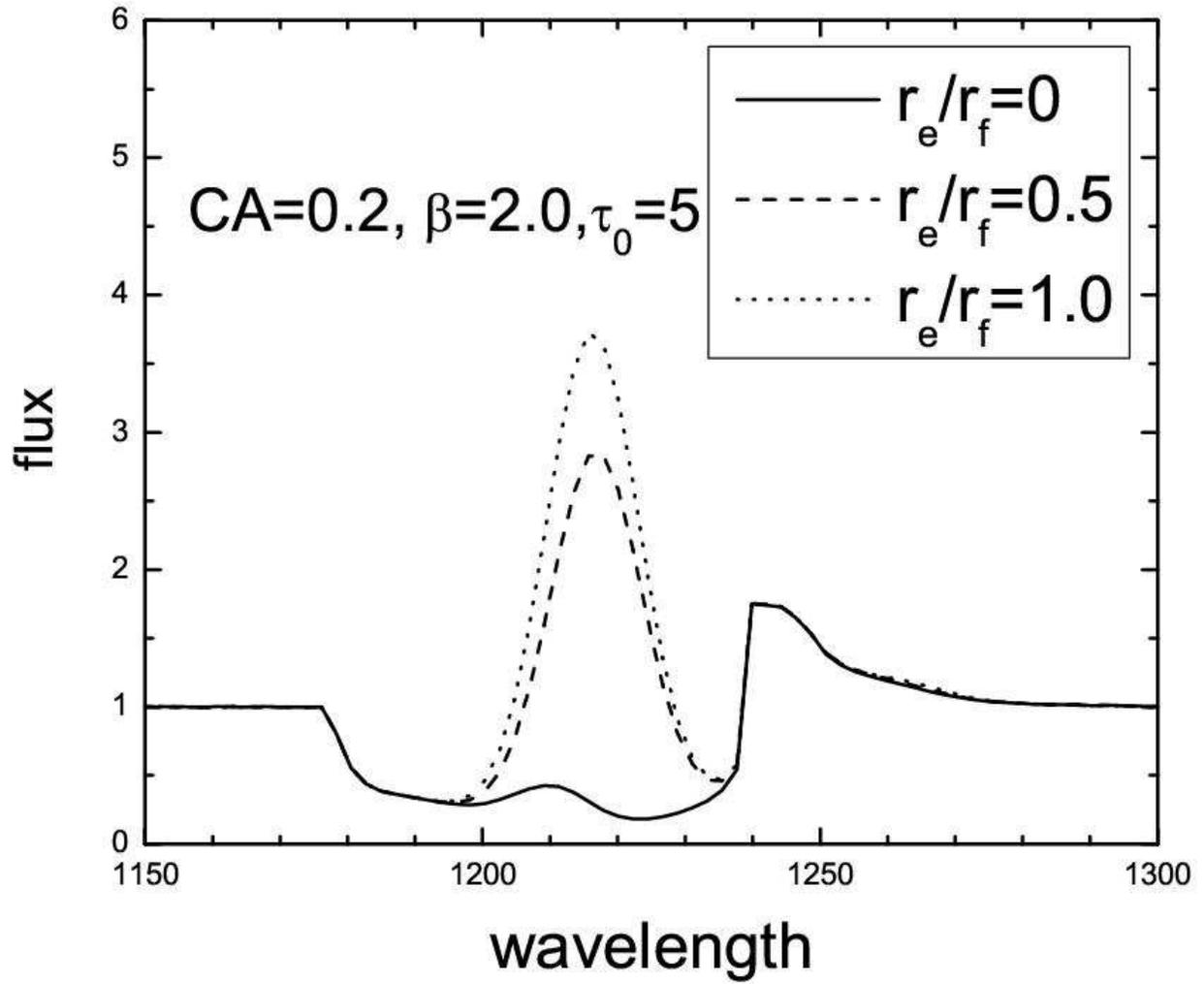} \caption{\label{fig_bal2} Mean spectra of
BAL QSOs for three models with parameters shown in the panel.}
\end{figure}

To explore the role played by the relative size of the Ly$\alpha$
emitting region in determining the scattered flux, we perform
simulations with $r_e/r_f$=0 and 1 (Model 3 and 4, respectively).
The model parameters and corresponding $\overline{{\rm EW}}_{\rm
ab}$ are listed in Tab. \ref{tab_ew}, and the average BAL spectra
are shown in Fig. \ref{fig_bal2}. The total scattered flux
increases only weakly with the size of the Ly$\alpha$ emission
region. When $r_e/r_f=0$, most of the Ly$\alpha$ photons along the
BAL direction are absorbed (solid line in Fig. \ref{fig_bal2}). It
is similar to the extreme case of HK96, in which a point-like BELR
and a large optical depth are assumed. As expected, the absorption
depth of the Ly$\alpha$ BEL in the mean BAL spectrum decreases
dramatically when $r_e/r_f=1$. However, the amount of absorbed
line photons initially traveling toward the disk increases
rapidly, and compensate for the decrease of seed photons
propagating along the BAL direction as $r_e$ increases. The
competition between the two components eventually results in the
weak dependence shown here. We also check the dependence of the
scattered flux on the parameter $\beta$ (Model 5$\sim$7), and list
the results in Tab. \ref{tab_ew}. Only a weak trend is found. Then
we carry out simulations with various mean optical depths, ranging
from 0.5 to 10. The model parameters, as well as the corresponding
$\overline{{\rm EW}}_{\rm ab}$, are also listed in Tab.
\ref{tab_ew} (Model 8$\sim$11). The total scattered flux increases
markedly with increasing optical depth up to $\tau_0\leq5$. At
$\tau_0=0.5$ the derived absorption EWs are very similar to the
average-A case in HK96. The scattered emission gets stronger for a
larger optical depth, at least for $\tau_0\geq1.5$.

The scattered flux is sensitive to the optical depth of the
outflow, besides its covering factor and the EW of the Ly$\alpha$
emission. This explains why our favored model can yield a larger
scattered flux than HK96, which adopted a smaller apparent optical
depth profile and a smaller relative size of the BELR (point-like)
than ours. Both models can reproduce the moderate absorption of
Ly$\alpha$ photons along the outflow direction as observed in the
spectra of BAL QSOs, but yield very different amounts of the
scattering emission. So additional observational constraints are
required to distinguish these two models. As mentioned in SEC.
\ref{sec_model}, a large optical depth in the BAL outflow is
suggested by recent studies (e.g., Hamann 1998; Arav et al. 1999a;
Wang et al. 1999; Lu et al. 2008). Moreover, it is found that  the
BELR and the outflow are comparable in size from spectroscopy
(Turnshek et al. 1988)
  and spectropolarimetry observations
(Goodrich \& Miller 1995; Ogle 1997; Ogle et al. 1999; Lamy \&
Hutsem$\grave{e}$kers 2004; Young et al. 2007).
In particular, in one
BAL QSO, 1603+3002, the BEL is not absorbed, implying a
large size of its BELR (Arav et al. 1999b).

\begin{figure}
\plotone{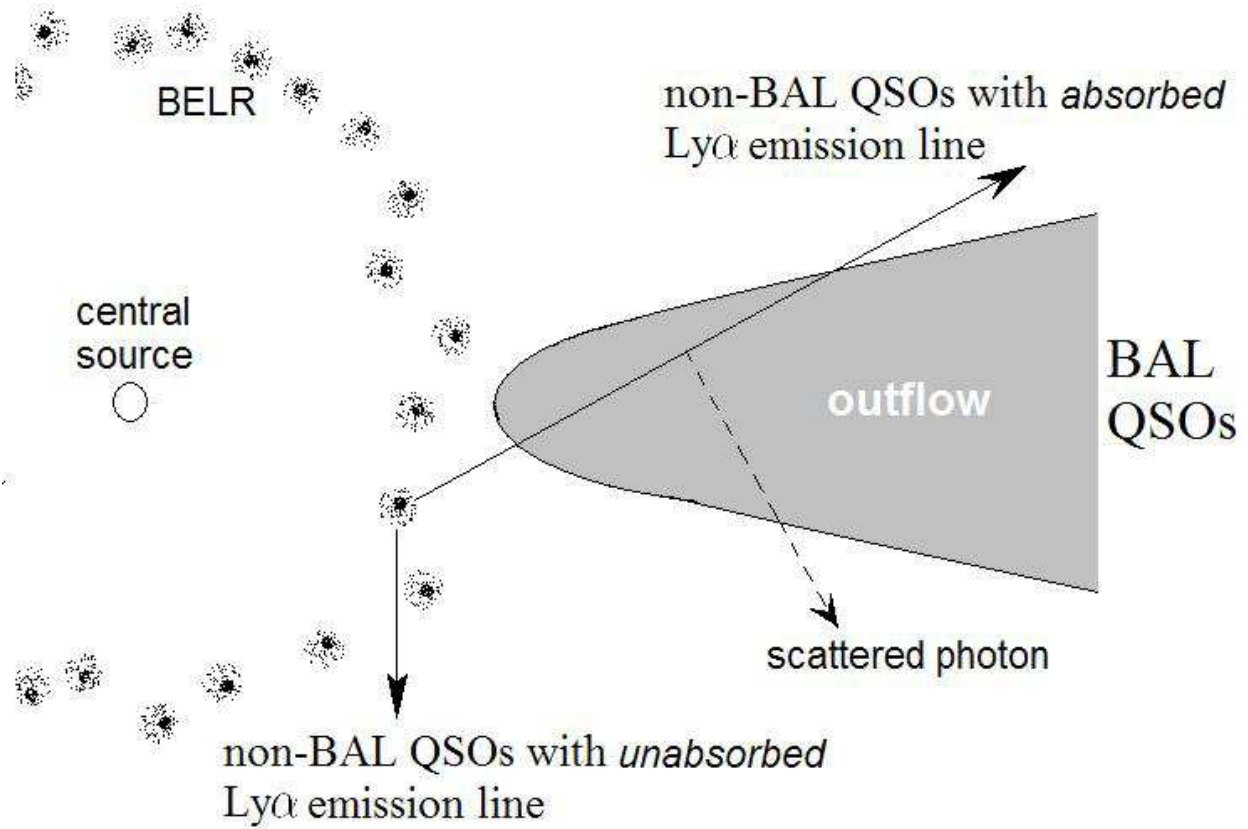}
 \caption{\label{fig_hkm} Cartoon plot for the geometry of equatorial outflow used by HKM. As the size of the BELR is comparable with
 the outflow, even if viewing along a BAL-free direction, one may also find that part of the BELR is blocked by the outflow.}
\end{figure}

The outflow can scatter Ly$\alpha$ photons  traveling initially
along BAL-free directions as long as  the Ly$\alpha$ emission
region has a size comparable to that of the outflow, regardless of
detailed models. In our model, Ly$\alpha$ photons traveling toward
the disk plane provide additional seed photons. If the absorption
by the accretion disk is negligible, such as the model proposed by
HKM, the situation would be slightly different. In Fig.
\ref{fig_hkm}, we illustrate with a sketch how the process works
in such a model. When viewing along some special BAL-free
directions as indicated in the figure, the Ly$\alpha$ emission is
partially blocked and scattered by the outflow (see Fig.
\ref{fig_hkm} for demonstration). In addition, this model predicts
a population of non-BAL QSOs that would display an absorbed
Ly$\alpha$ emission line. A candidate of such QSO was reported by
Hall et al.(2004), which exhibits strongly absorbed Ly$\alpha$
emission but an unabsorbed continuum. Interestingly, its \civ\
emission line shows significant blueshifted asymmetry, indicative
of the presence of an underlying outflow. Because only emission
lines are absorbed and the absorption line profile is dependent on
the detailed structure of both the outflow and the BELR, this type
of objects would be quite difficult to be recognized. This may
explain why only one such object has been reported to date (see
also Green (2006) for another possible case). Near-infrared
observation of Balmer lines can test the scenario for Ly$\alpha$
absorbed non-BAL QSOs since Balmer lines are not expected to be
absorbed by the BAL gas. It is worth noting that such a population
of non-BAL QSOs are also expected in the model in which absorption
by the accretion disk is considered, for example the BALR and BELR
geometry proposed by Elvis (2000).

\section{Profile of Scattering Emission}\label{sec_pro}

The \nv\ BEL is generally severely blended with the stronger
Ly$\alpha$ line, and its EW is usually measured by using some
de-blending technique. Therefore, the measured EW of \nv\ line is
also dependent on the technique used. Usually, a narrower
scattering profile can contribute more to the measured EW. The
quantity $f_c={\rm EW_{HK}/EW_{sc}}$ defined above can be used to
describe approximately the width of the scattering line. The mean
$f_c$ in BAL-free directions is 0.79 (0.81) for Model 1 (2), which
is larger than HK96 ($\leq0.18/0.31\simeq0.58$). Our models
produce obviously narrower profiles than HK96. In this section we
investigate the several processes that affect the width of
scattered line. Moreover, we will show that the relative narrow
profile is a natural result of our model.

\subsection{Radial Velocity Structure and Optical Depth}

\begin{figure}
\plotone{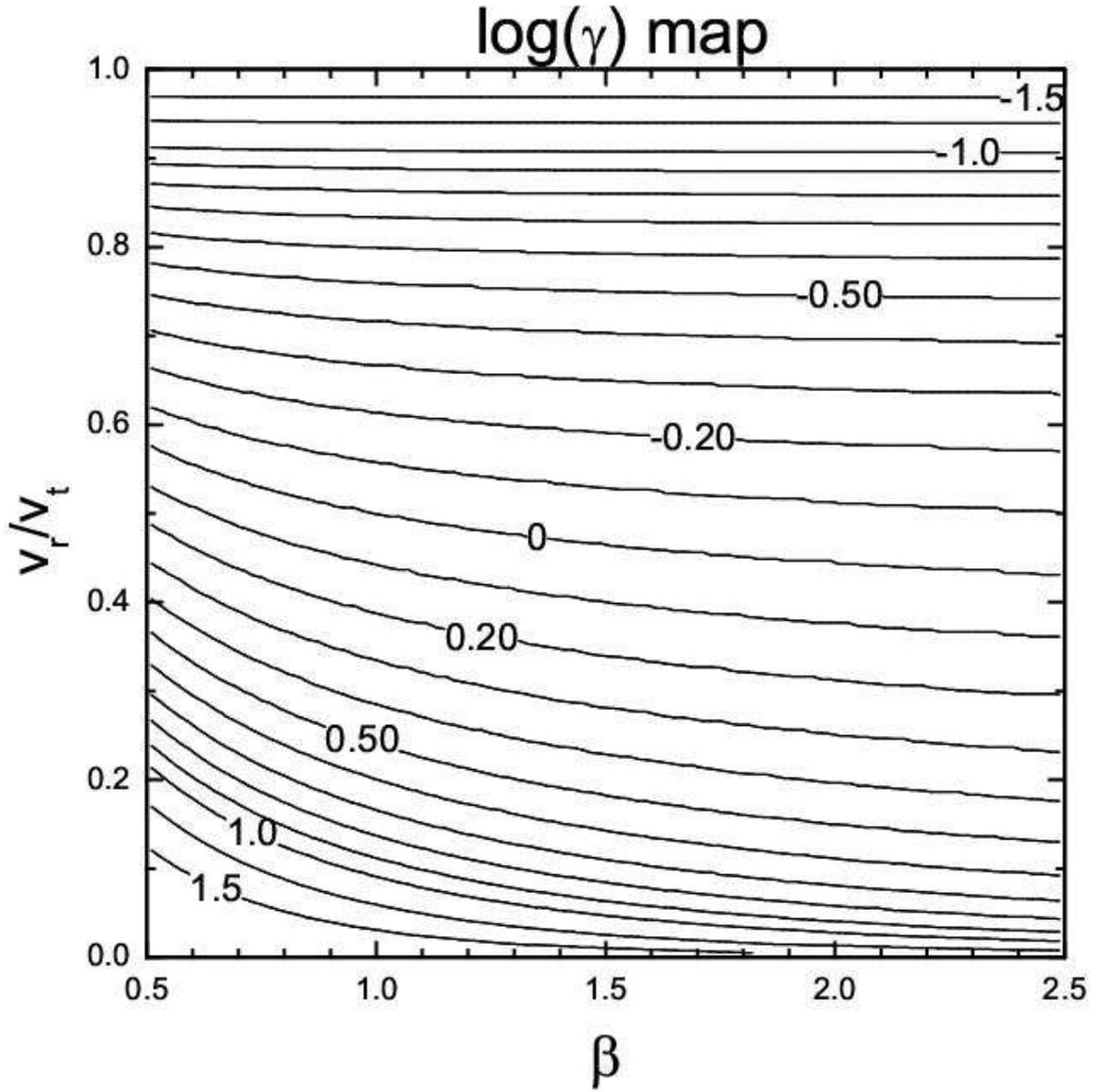}
 \caption{\label{fig_gamma} $\log\gamma$, where $\gamma$ is the ratio of the optical depth along the direction perpendicular to the radial direction
 to the optical depth along the radial direction, as functions of $\beta$ parameter and scaled velocity $v_r/v_t$.}
\end{figure}

For a given geometry the scattering line profile depends on the
angular distribution of the optical depth (HKM). In the Sobolev
approximation, the optical depth along an arbitrary direction $z$
can be written as
\begin{equation}
\tau_z=\tau_r(v_r)|\frac{dv_r}{dr}|/|\frac{dv_z}{dz}|
\end{equation}
where
$v_r$ and $v_z$ are the flow velocity in the radial and $z$
directions, respectively. To describe the anisotropic distribution
of the optical depth, we define a parameter $\gamma=\tau_p/\tau_r$,
where direction $p$ is perpendicular to the radial direction (HKM, HK96).
$\gamma$ can be expressed as
\begin{equation}
\gamma=\frac{\tau_p}{\tau_r}=\frac{|dv_r/dr|}{|dv_p/dp|}=\beta[(\frac{v_t}{v_r})^{1/\beta}-1]
\end{equation}
The second equation is valid only for our radial velocity profile.
The scattering is isotropic for $\gamma=1$, radial for
$\gamma\gg1$ and tangential (perpendicular to radial) for
$\gamma\ll1$. In Fig. \ref{fig_gamma}, we show $\log\gamma$ as
functions of $\beta$ and the scaled radial velocity $v_r/v_t$.
$\gamma$ is insensitive to $\beta$ in the range considered, and
rapidly decreases with increasing radial velocity. In most of the
radial velocity range, $\gamma$ varies from $10^{-1.5}$ to
$10^{1.5}$. The dynamic range in the $\gamma$ parameter is broader
than that considered by HK96 and HKM, from 0.2 to 20. It is
interesting that photons tend to be scattered into the radial
direction in the inner region of the flow (low $v_r/v_t$), and
along tangential directions in the outer part of the flow. This
means that the velocity structure in M95 tends to generate a
relatively narrow scattering profile.

\begin{figure}
\plotone{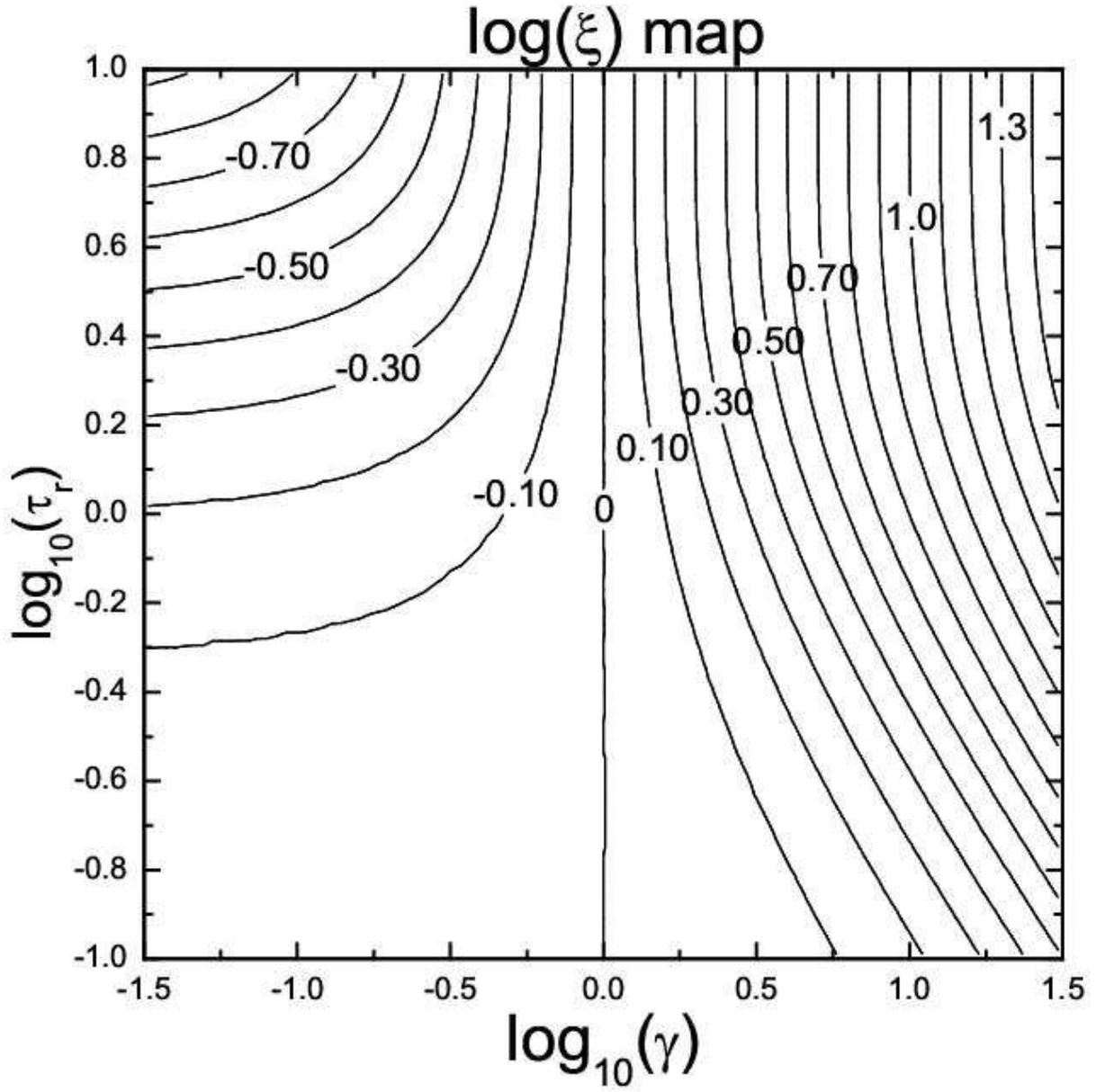}
 \caption{\label{fig_ep} $\log \xi$ (see the definition of $\xi$ in the text)
 as functions of radial optical depth $\tau_r$ and $\gamma$
parameter.}
\end{figure}

The $\gamma$ parameter gives only  a {\em qualitative}
description of anisotropic scattering. According to
Castor(1970), the probability of a photon escaping from a given point
along the  $z$ direction is
\begin{equation}
\kappa_z=\frac{1}{\kappa'}\frac{1-e^{-\tau_z}}{\tau_z}
\end{equation}
where $\kappa'$ is the normalized factor. To describe the
anisotropic distribution of the escape probability, we introduce a
parameter
\begin{equation}
\xi=\frac{\kappa_r}{\kappa_p}=\gamma\frac{1-e^{-\tau_r}}{1-e^{-\gamma\tau_r}}.
\end{equation}
We show $\log \xi$ as functions of $\tau_r$ and $\gamma$ in Fig.
\ref{fig_ep}. The probability distribution of the escaping
direction is related to both $\gamma$ and optical depth. The
tangential scattering, which prefers to generate a narrow
scattering line profile, becomes effective only when $\tau_r$ is
large enough. HK96 used an apparent optical depth structure,
derived from observed BAL spectra, to calculate the scattering
profile. However, this apparent optical depth method severely
underestimates the real optical depth (Arav et al. 1999a;1999b).
It may explain partially why the HK96 model produces a broader
profile than ours.

\subsection{Doublet Transitions}%\label{sec_dt}

\begin{figure}
\plotone{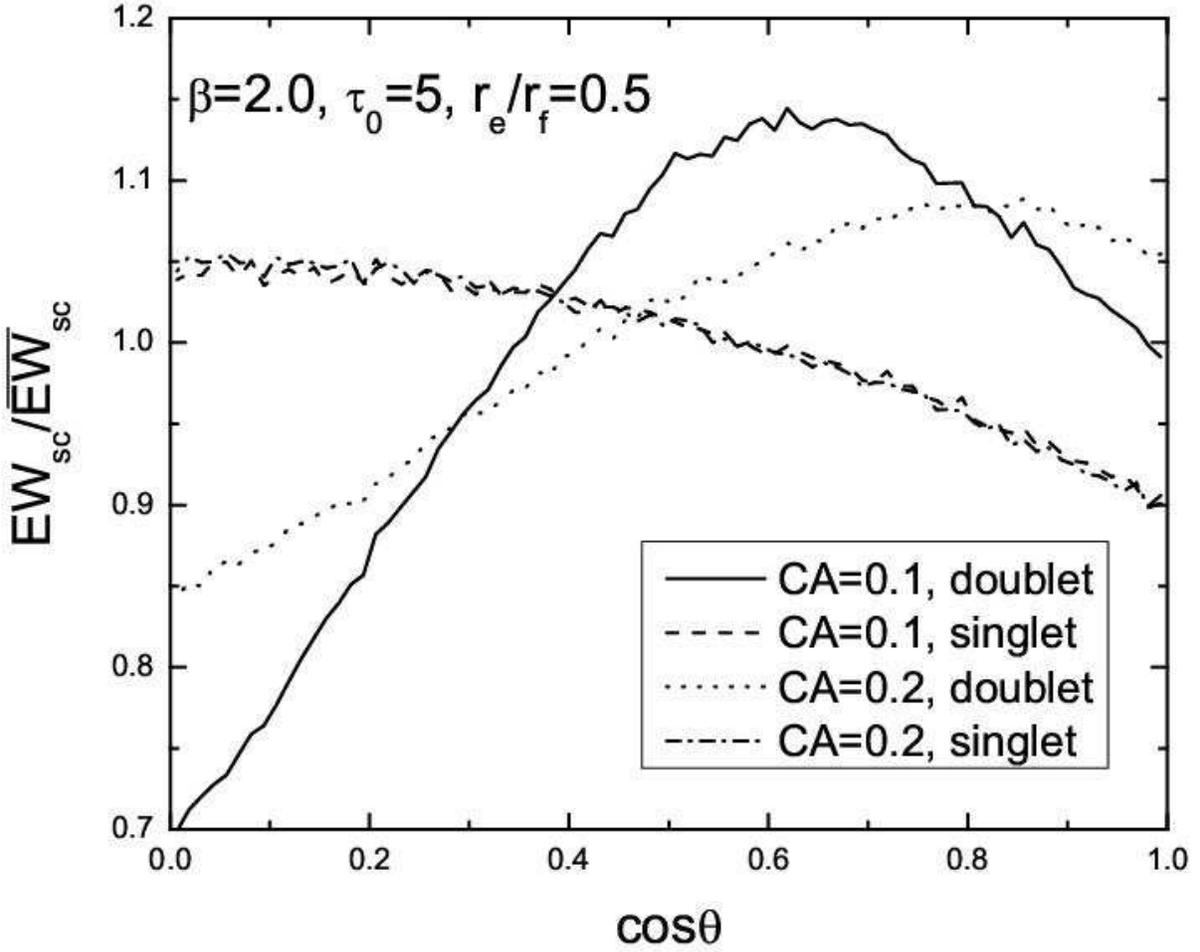}
 \caption{\label{fig_ds} Normalized EW of the total scattering emission, i.e. EW$_{\rm sc}$/${\rm\overline{EW}_{sc}}$,
 as a function of $\cos\theta$ for both singlet and doublet transitions. The model parameters are shown in the panel.}
\end{figure}

The escape probability method was developed to approximate the
singlet resonant scattering. However,
\nv$\lambda\lambda1238.8,\;1242.8$\A\ is a doublet line, in which
the fine structure splitting is as broad as about 970~\kms. We
refer to the transition associated with shorter (longer)
wavelength as T$_1$(T$_2$). This large velocity separation between
the two transitions would induce a quite different radiation
transfer effect from the singlet transition in a non-spherical
outflow. In Fig. \ref{fig_ds} we show the EW of the total
scattering emission line as a function of inclination for both
doublet and singlet resonant scattering. To compare results using
different CA values, the EWs are shown as normalized to each ${\rm
\overline{EW}_{sc}}$.

Two things are worth-noting in this figure. First, for singlet
transition, the scattered flux  increases monotonously as the
angle of the line-of-sight to the equatorial plane decreases.
However, in the case of doublet transition, the scattered flux is
suppressed along the equatorial direction. Considering a photon
emitted from the central source and incident upon the outflow, it
first encounters the scattering surface associated with transition
T$_1$. After one or more scatters within a tiny volume of size
corresponding to several thermal widths (hereafter T$_1$ volume),
the photon emerges along a new direction $\textbf{d}$ (Lee \&
Blandford 1997). If the outflow is  accelerated monotonously, the
fluid is expanding in all directions around the T$_1$ volume.
Therefore, in a spherical outflow, the photon will encounter
transition T$_2$ regardless of the emergent direction
$\textbf{d}$; that is to say, the T$_2$ scattering surface
encloses the corresponding T$_1$ volume. When a  non-spherical
geometry is considered, for example an equatorial outflow, the
situation is different. Since an equatorial outflow is thin along
the polar direction, the polar area of the T$_2$ surface may lie
outside the outflow. If $\textbf{d}$ is nearly polar, there is a
chance for the  photon to escape from the outflow without
experiencing T$_2$ scattering. Otherwise, it may encounter
transition T$_2$ and be scattered again. This process results in a
suppression of the scattered emission along the equatorial
direction. On the other hand, the scattered light along the polar
direction is dominated by photons experiencing T$_1$ scattering
only, so the EW along the polar direction appears to be similar to
the singlet case, i.e., decreasing with increasing $\cos\theta$.
This yields a turnover in the EW$_{\rm sc}\sim\cos\theta$
relationship.

At which orientation the turnover point occurs depends on the
opening angle of the polar area of the T$_2$ surface, which lies
outside the outflow, with respect to the T$_1$ volume. This
opening angle is related to the shape and size of the T$_2$
surface and the covering factor of the outflow. As the covering
factor decreases, the fraction of the T$_2$ surface within the
outflow becomes smaller, and then the turnover point moves toward
the equatorial direction. This is the second noteworthy point in
Fig. \ref{fig_ds}. We note that the shape and size of the T$_2$
surface are related to the velocity splitting between the two
lines and the velocity structure of the outflow, and thus the
dependence of EW$_{\rm sc}$ on $\cos\theta$ is also relevant to
these two properties. Here we only give a qualitative analysis
about this effect; a detailed interpretation is beyond the scope
of this paper. For a comparison, the results for singlet are
independent of CA (Fig. \ref{fig_ds}).

Doublet resonant transitions tend to scatter photons into the
polar direction. Since the width of the scattering line profile is
the narrowest along the polar direction, this process reduces the
width of the scattered flux on average.

\section{Observational Evidence for Scattering Emission}\label{sec_evi}

The difference between the mean spectra of BAL and non-BAL QSOs
can be used to constrain the contribution of the scattering
emission (Weymann et al. 1991; HKM). We have shown that the
scattering emission along the BAL direction is mildly smaller than
that along BAL-free directions (Fig. \ref{fig_ew}, see also SEC.
\ref{sec_enh}). However, if there is a wide distribution in the
covering factor, BAL QSOs have a larger mean covering factor than
non-BAL QSOs due to selection effects (Morris 1988; Wang et al.
2005). Since the scattered flux increases with the covering
factor, the average EW$_{\rm sc}$ for BAL QSOs can be larger than
that for non-BAL QSOs. For instance, assuming a mean value of CA
of 0.2 for BAL QSOs, the resultant EW$_{\rm sc}$ is 12.1\A\ for
BAL QSOs based on our model, while it is only 7.3\A\ for non-BAL
QSOs, if the mean CA in non-BAL QSOs is 0.1.

Weymann et al.(1991) found that there appears to be an excess of
\nv\ in the difference spectra of BAL and non-BAL QSOs. A close
inspection indicates that the excess flux  extends up to ten
thousand \kms\ redward of the \nv\ line center (see figure 3 in
HKM for a more clear illustration). The excess is much broader
than the emission from the normal BELR, whereas its width is
similar to the typical velocity of the BAL outflows. Therefore
this is direct evidence for a scattering line. We note that the
amount and width of the excess flux may be larger than what one
obtains directly from the difference spectrum, since part of the
photons scattered by N$^{+4}$ ions fill in the \siiv\ BAL. In
Weymann's sample, the \civ, \nv\ and \siiv\ BALs extend  up to on
average 25000\kms\ blueward of the relevant emission lines. There
is a narrow core in the excess flux, which might result from
strong tangential scattering in (at least part of) the BALR, or
from different intrinsic emission lines in two kinds of QSOs.

If there is a large contribution from resonant scattering in the
BALR, the \nv\ emission would exhibit a polarized feature very
different from other lines produced in the normal BELR. Normal
BELs, such as \civ, \siiv\ and \mgii, are usually absent in
polarized spectra (Goodrich \& Miller 1995; Schmidt \& Hines 1999;
Ogle et al. 1999; Lamy \& Hutsem$\grave{e}$kers 2004). In contrast
to these lines, unusually strong and broad  emission of \nv\ in
polarized spectra were found by Lamy \&
Hutsem$\grave{e}$kers (2004) in all of their BAL QSOs spectra with
available \nv. This prominent feature is also present in the
polarized spectra of several BAL QSOs from the spectropolarimetric
atlas of Ogle et al.(1999). This feature can be interpreted as a
result of the resonant scattering of Ly$\alpha$ photons in a
non-spherical outflow.

\begin{figure}
\plotone{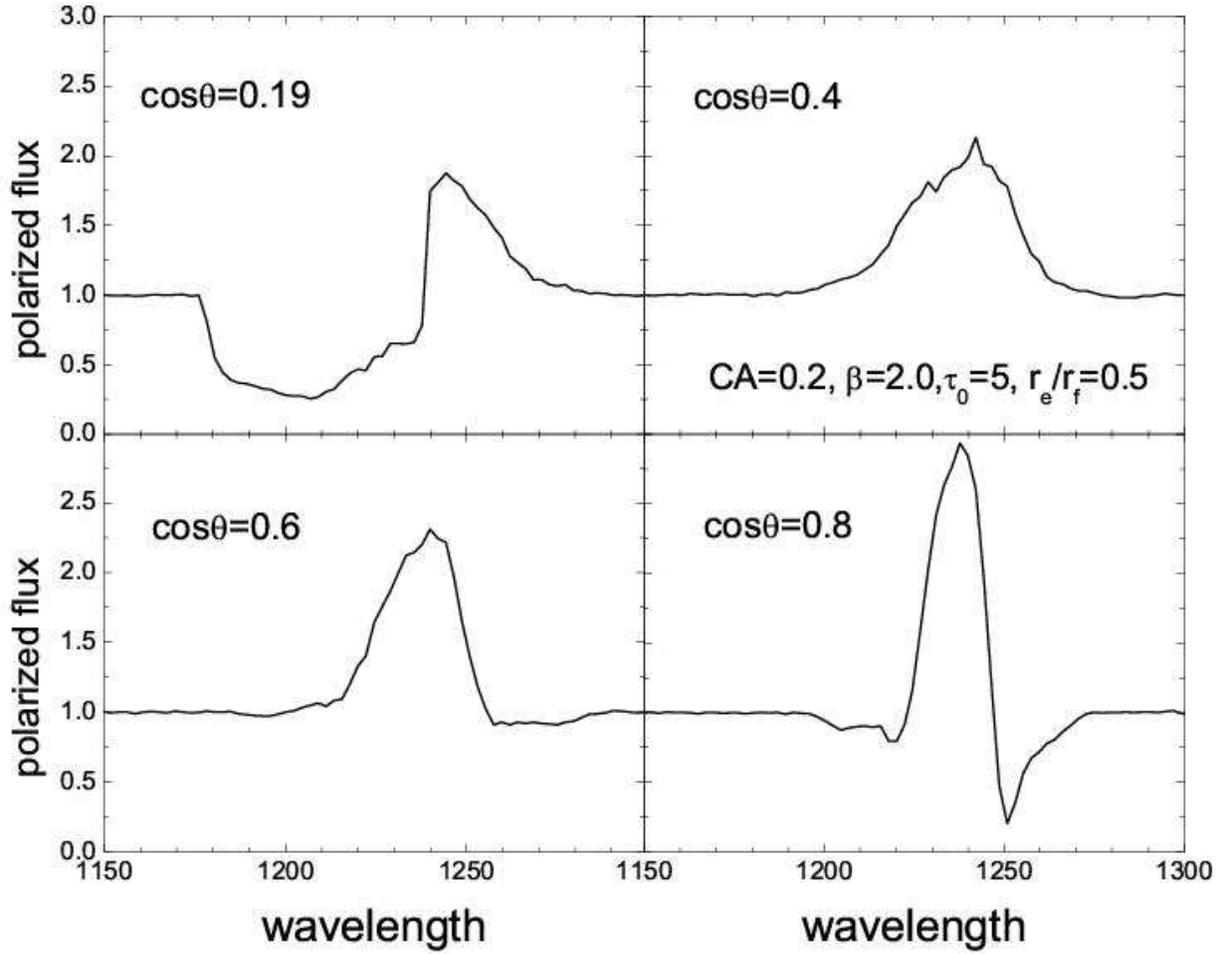}
 \caption{\label{fig_pf} The polarized flux as a function of wavelength in four different directions denoted in each panel.
 The model parameters are shown in the top-right
panel. The polarized spectra are normalized to the continuum
polarized flux.}
\end{figure}

In Fig. \ref{fig_pf}, we show the polarized spectra for Model 2 in
one BAL direction and three BAL-free directions. The spectra are
normalized to the polarized continuum. In order to compare our
results with observations, we set the polarization degree of
incident continuum and intrinsic emission lines to 1.5\% and zero
respectively, which are consistent with observed polarization
(e.g., Lamy \& Hutsem$\grave{e}$kers 2004). We note that the EWs,
profiles and especially the polarized flux of the resonantly
scattered photons are independent of the polarization of incident
light because the incident polarization is generally far less than
100\%. As one can see, resonant scattering yields prominent
emission feature around the \nv\ line in the polarized spectra of
both BAL and non-BAL QSOs. However, these simulated polarized
features are weaker compared to the observed one in Lamy \&
Hutsem$\grave{e}$kers (2004). It implies that the covering factor
or/and the EW of the Ly$\alpha$ emission line of these QSOs may be
larger than the values we adopted, or the outflow is
non-axisymmetric with respect to the rotation axis of the
accretion disk, or the incident continuum and line emission are
anisotropic. The unusually large polarized emission provides
important and strong evidence for that \nv\ is significantly
enhanced by resonant scattering  in these BAL QSOs. Suppose  BAL
and non-BAL QSOs are from the same parent population, one would
naturally expect strong scattered emission in the polarized as
well as the total spectra of non-BAL QSOs (Fig. \ref{fig_f10},
\ref{fig_f20} and \ref{fig_pf}). Extensive spectropolarimetric
observations in this waveband were focused on BAL QSOs, thus
high-resolution, polarized spectra of non-BAL QSOs are certainly
required. Future spectropolarimetric observations will provide us
with a unique insight for understanding the BALR properties and
scattering enhancement.

The reverberation mapping technique offers another way to
distinguish the two different origins of the \nv\ emission. If
most of the \nv\ BELs can be attributed to the scattering in the
BALR, the size of the \nv\ BELR will be larger than that of
Ly$\alpha$. Otherwise, it should be smaller than that of
Ly$\alpha$ and of other lower-ionization lines, since
high-ionization lines are formed closer to the continuum source
according to previous reverberation studies. This can be tested
with reverberation mapping observations of \nv. Previous
reverberation mapping studies of Seyfert galaxies in UV suggested
that the \nv\ emission line region is smaller than that of
Ly$\alpha$, \civ, H$\beta$, etc (e.g., Reichert et al. 1994;
Korista et al. 1995; Wanders et al. 1997; Rodriguez-Pascual et al.
1997; O'Brien et al. 1998). However, resonant scattering is
expected to be small in those low-luminosity objects because of
the low velocities of their outflows of the order of hundreds
\kms, which are insufficient to scatter a substantial fraction of
Ly$\alpha$ photons. Meaningful tests can only be done for luminous
QSOs, in which the BAL outflow is believed to be launched.
Unfortunately, monitoring of QSOs UV spectra has only just been
started, and no firm conclusion has been reached so far. Recently,
Kaspi et al. (2007) carried out spectrophotometric monitoring of
several luminous QSOs at redshifts z = 2.2-3.2, for which \nv\
enters the observing window. However, the \nv\ and Ly$\alpha$
lines could not be deblended due to the low spectral resolution.
Future QSOs monitoring with medium spectroscopic resolutions may
provide crucial tests of scattering enhancement. Furthermore, this
technique will also provide useful probes to the BAL outflows, if
scattered Ly$\alpha$ photons indeed contribute significantly to
the \nv\ emission.

\section{Origin of the Peculiar Behavior of \nv\ BEL}\label{sec_nv}

Flux ratios between \nv\ and other lines, such as \ovi, \civ\ and
\heii, are often used to determine the metallicities of the BEL
gas (e.g., Hamann \& Ferland 1993;1999, Hamann et al. 2002).
However, there exists a systematical discrepancy between the
metallicities inferred by the line ratios using \nv\ and using
some weak semi-forbidden lines (Shemmer \& Netzer 2002; Dietrich
et al. 2003, hereafter D03). Shemmer \& Netzer investigated a
large sample of AGNs and found that the metallicities derived from
\niv/\civ\ are systematically smaller than those obtained by using
\nv/\civ\ by a factor of 3 or 4. This discrepancy was confirmed by
an immediate follow-up study (D03). Meanwhile, D03  found that the
mean \nv/\civ\ metallicity is larger by a factor of $\sim$1.7 than
those obtained using \niii/\oiii.

It has been suggested that these  discrepancies may arise from the
unsuitability of using these weak lines: the \niv/\civ\ line ratio
is sensitive to the radiative transfer effect, which is ambiguous
(D03); and \niii/\oiii\ is related to the distribution of the
column density of clouds and may yield merely lower limits on the
metallicity in the worst case (Hamann et al. 2002). If it is the
case, suppose the line ratios involving \nv\ are reliable
metallicity indicators, one would expect the distribution of the
\nv\ metallicity might be narrower than those inferred using the
two weak semi-forbidden lines ratios. However, a much larger
dispersion in the \nv\ metallicity than those based on the weak
lines is clearly presented in D03 (their figure 3 and figure 5a).
It implies that the \nv\ emission, rather than those
semi-forbidden lines, is affected by some other processes. The
resonant scattering proposed in this paper would play a role, as
it can contribute significantly to the \nv\ emission based on our
model. For a covering factor of 0.2, the \nv\ emission may be
enhanced on average by a factor of 1.82  (using HK96's EW data) or
2.73 (using our supposed data). These values are broadly
consistent with the observed difference between the metallicities
inferred from \nv\ and  from some of the weak semi-forbidden lines
ratios. In particular, the strength of the scattering emission has
a wide distribution, which naturally explains the large scatter of
the `\nv\ metallicity'.

In order to explain the absence of the Baldwin effect in the \nv\
line, Korista et al. (1998) invoked a strong positive correlation
between metallicity and luminosity in their `locally optimally
emitting cloud' model. Difficulties for such an interpretation
are: 1) the strong correlation between metallicity and luminosity
is obtained by using the \nv\ emission line (e.g., Hamann \&
Ferland 1993; Shemmer \& Netzer 2002), while the metallicity
inferred by the semi-forbidden lines  shows only a weak dependence
on luminosity (D03); 2) this model overestimates severely the EW
of other two nitrogen lines, \niv\ and \niii, at the high
luminosity end (Dietrich et al. 2002).

The peculiar behavior of \nv\ can be understood in terms of
scattering emission. There is growing evidence that the BAL
phenomenon is correlated with some of the fundamental properties
of QSOs. Boroson (2002) has shown that BAL QSOs tend to occupy the
extreme end of the Boroson \& Green (1992) Eigenvector 1, which is
thought to be driven by the Eddington ratio. The absorption
strength and maximal velocity of BALs are found to correlate with
luminosity and the ratio between the intrinsic X-ray and optical
luminosities (Laor \& Brandt 2002; Ganguly et al. 2007; Fan et al.
2009). Ganguly et al. (2007) found significant correlations
between the fraction of BAL QSOs, or the average covering factor
of the outflow, and luminosity/Eddington ratio/black hole mass.
Therefore, the scattering emission increases with QSO luminosity
and hence compensates for the loss of the intrinsic \nv\ EW due to
the normal Baldwin effect. A detailed calculation requires the
dependence of properties of QSO outflow on luminosity, which,
however, suffers from selection biases and uncertainties caused by
QSO identification methods.

\section{Funnel-like Outflows}\label{sec_fun}

In contrast  to the simple model in Fig \ref{fig_dw}, a
complication comes from the presence of a postulated  dusty torus,
which is invoked for the unification model of the two types of
Seyfert galaxies. The evidence for such tori in active galactic
nuclei is compelling, including a large population of type 2 QSOs
suggested by recent optical and infrared observations (e.g.,
Zakamska et al. 2003; Stern et al. 2005). The accretion disk
likely has the same symmetry axis as the dusty torus since it is
generally believed that the disk is fed by the gas from the torus.
In such a configuration, an equatorial disk-like outflow will
encounter an optically-thick dusty torus, and a strong interaction
between the two is expected. Unless  the outflow can blow away the
torus, it will be blocked by the latter. One possibility is that
the BAL gas arises initially nearly vertically, and is then
accelerated radially by radiation pressure in a funnel-like
region, as proposed by Elvis (2000; see also Young et al. 2007).
In such a model, the vertical outflow will be located inside the
torus, whereas the radial outflow moves outward freely. If the
boundary of the torus is set by the interaction of the torus and
radial outflow, the subtending angle of the torus would be similar
to that of the vertical outflow.

The funnel-like outflow with a dusty torus can have two major
consequences on the scattering emission. First, the covering
factor of the BALR would have been over-estimated using the
fraction of BAL QSOs. Because the vertical part of an outflow has
a small radial velocity gradient, and is not important for the
scattering enhancement. Thus the covering factor of a radial
outflow can be approximated by $N_{\rm BAL}/(N_1+N_2)$, where
$N_{\rm BAL}$, $N_1$ and $N_2$ are the numbers of BAL QSOs, Type 1
and Type 2 QSOs, respectively. Taking $N_2=N_1$ (e.g., Reyes et
al. 2008), the covering factor is only half of the detection
frequency of BAL QSOs among optical selected QSOs. The effect of
this on the scattering emission can be quantitatively estimated
via Eq. \ref{eq_calew}. Second, the incident angle relative to the
outflow surface has  increased significantly compared to the case
of an equatorial outflow. The radiative transfer effect results in
a different angle-dependence in the scattering flux.

\begin{figure}
\plotone{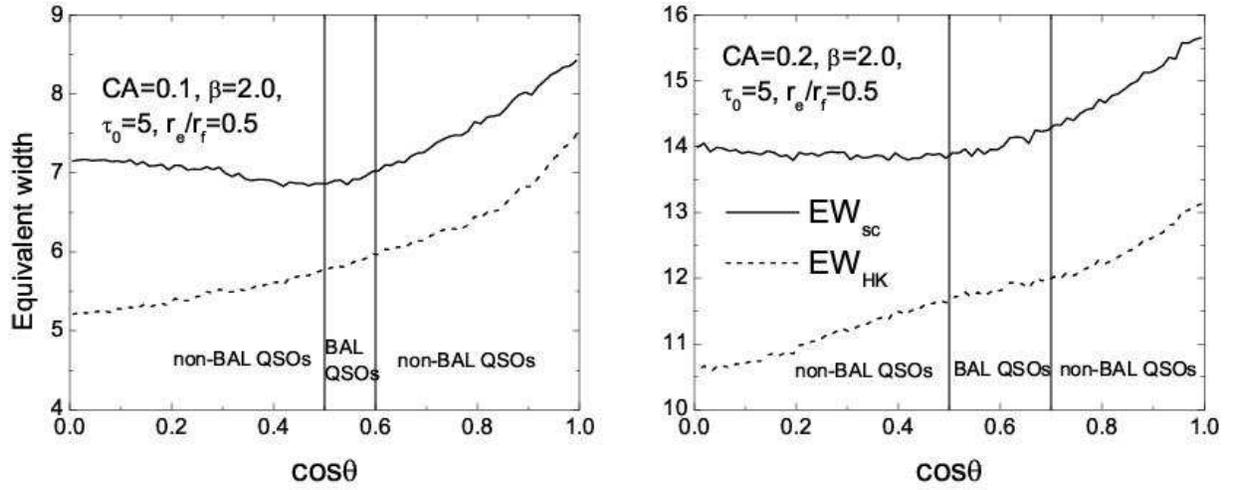}
 \caption{\label{fig_ewb} The EW, EW$_{\rm sc}$ and EW$_{\rm HK}$,
 as a function of $\cos\theta$ for funnel-like geometric models. The two vertical lines in each panel
 denote the boundary between BAL and non-BAL QSOs. The model
parameters are shown in the panels.}
\end{figure}

To assess quantitatively the latter effect, we perform two
simulations for a funnel-like outflow, similar to Model-B in WWW
but with Ly$\alpha$ scattering included. As in SEC.
\ref{sec_model}, BELs are assumed to be produced in a spherical
shell, and the intrinsic EW of Ly$\alpha$ and \nv\ are, 80 and
10\A, respectively. In one simulation, the outflow (referred to as
Model B1) has a covering factor of 0.1. Its upper and lower
surfaces are 53.1\deg~and 60\deg\ to the rotational axis,
respectively. In the second simulation, the inclination angles for
the upper and lower surfaces are 45.6\deg~and 60\deg (referred to
Model B2), respectively. The model B has a larger covering factor
of 0.2. The EWs of scattering emission, ${\rm EW_{sc}}$ and ${\rm
EW_{HK}}$, are shown in Fig. \ref{fig_ewb}. Other model parameters
are marked in each panel.

The mean EW$_{\rm sc}$ in BAL-free directions is 7.3\A\ (14.2\A)
for Model B1 (B2), the same as that of Model 1 (2), whereas mean
EW$_{\rm HK}$ is about 5.9\A\ (11.6\A), slightly larger than that
for the corresponding equatorial model (Tab. \ref{tab_ew}). These
results indicate that the average scattering enhancement of the
\nv\ emission is insensitive to the detailed geometry of the
outflow. It is simply because the optical depth and covering
factor of the BAL outflow and the intrinsic EWs of emission lines,
which determine the amount of the scattered photons, are the same
for Model B1 (B2) and Model 1(2). However, in the funnel-like
models, EW$_{\rm sc}$ decreases first, then increases with
$\cos\theta$, resulting in a minimum roughly at the BAL
directions, which is very different from that in the equatorial
model shown in Fig. \ref{fig_ew}. Such a EW$_{\rm sc}$
distribution can also be understood in terms of the combination of
doublet transitions plus the outflow geometry. As discussed in
SEC. 5.2%\ref{sec_dt},
,doublet transitions tend to suppress the
scattering emission along the directions in which the outflow is
thick enough.

Up to now, we ignore the scattering on the base of the biconical
outflow. This is valid if the vertical outflow has very high
ionization so N$^{+4}$ ions are very rare or the radial velocity
at the flow base is very low. In the latter case, given the large
covering factor of vertical outflows, the scattering of continuum
photons from that part of the flow may produce a prominent
relative narrow line in the absence of rotation. If the flow is
rotating sufficiently fast, it can resonantly scatter Ly$\alpha$
as well, and a narrow feature can be rather prominent in a face-on
system. Detailed calculations need to take into account both the
kinematics and geometry of the Ly$\alpha$ emitting gas, and is
beyond the scope of this paper. Note also that the radial velocity
in the shielded outflow can be substantial. Smooth curvature of
flow lines are suggested in a magnetocentrifugal driven outflow
model (Blandford \& Payne 1982; Konigl \& Kartje 1994; Everett
2005; Miller et al. 2006). In that case, the flow base possibly
spans a relatively large velocity range so that a substantial
fraction of Ly$\alpha$ photons may be scattered there. In that
case, the effective covering factor can be significantly larger
than the estimate above.

\section{Summary}\label{sec_sum}

The \nv\ BEL shows a different behavior from other nitrogen lines,
\niv\ and \niii, and other high ionization lines, such as \civ\
and \ovi, indicative of a different or additional origin. It has
been suggested that resonant scattering in the BALR can contribute
significantly to the observed \nv\ emission line. In this paper,
we derive the profile and EW of the scattering line by using the
Sobolev Monte Carlo approximation. Some new observational
constraints on the geometry and optical depth of QSO outflows are
taken into account in our calculation.

By adopting a more realistic model than those in previous work, we
find that the observable \nv\ emission, EW$_{\rm HK}$, is
 increased by a factor of 1.82--2.73 on average. Our value is larger
than the previous estimates. This can be attributed to the larger
covering factor and radial optical depth we adopt than those used
in earlier work. We emphasize that the average covering factor and
optical depth adopted in our model are consistent with results
from recent observational studies. The outflow, albeit with a
large radial optical depth, can scatter only a small fraction of
Ly$\alpha$ photons along the BAL direction in our model, since the
size of the BELR is comparable to the launching radius of the
outflow. The relative sizes of the BELR and outflow adopted here
are well constrained from many spectropolarimetric data. Our
calculations find a wide range for the contribution of the
scattered photons. It depends on several parameters, such as the
covering factor, the optical depth, Ly$\alpha$ EW and the
orientation, which vary from one QSO to another. As a result, the
scattering emission may dominate the \nv\ emission in some QSOs,
whereas negligible in others.

Another reason why the  scattering contribution can be significant
is that our model can produce narrow scattering profiles. About
80\% of the total scattered flux contributes to the \nv\ BELs as
measured. A simple analysis indicates that the large optical depth
we adopt can make tangential scattering to become effective, and
thus yield a relatively narrow average profile. In addition, the
large splitting between the \nv\ double lines tend to enhance
(suppress) the scattering emission along the polar (equatorial)
direction in an accelerated disk-like outflow. Since the profile
of the scattering line along the polar direction is narrower than
that along the equatorial direction, doublet transition can
produce narrower profiles than singlet on average.

The \nv\ excess in the mean spectra of BAL QSOs compared to that
of non-BAL QSOs provides direct observational evidence for the
scattering contribution. The width of the excess flux is too broad
to be concordant with emission from the normal BELR, but is
similar to the typical velocities of BAL outflows. The unusually
large and broad \nv\ emission in polarized spectra of BAL QSOs is
another piece of strong evidence in support of our argument. Our
calculation suggests that this feature can be yielded via resonant
scattering of Ly$\alpha$ photons in the outflow. More importantly,
our models produce similar prominent features in the polarized
flux spectra of non-BAL QSOs. Future spectropolarimetric
observations of QSOs, especially non-BAL QSOs, may provide crucial
constraints on this issue, and further a useful probe of the
physical condition of outflow. Reverberation mapping technique
offers another way to investigate the importance of the
contribution of the scattered light. Since low-luminous AGNs
harbor only low-velocity outflows of the order of hundreds \kms,
which is insufficient to scatter a substantial fraction of
Ly$\alpha$ photons, this test can be done for luminous QSOs only,
in which the BAL phenomenon is observed.

Finally, we discuss the role of resonant scattering in
establishing the peculiarity of the \nv\ emission, i.e. the
absence of \nv\ Baldwin effect and the discrepancy between the
metallicities inferred by \nv\ and by \niv\ and \niii. We argue
that scattering in the BALR offers a more promising explanation to
the \nv\ peculiarity over those proposed in previous work.

\acknowledgments We thank the referee for a useful report. We
thank Ms. Chengkun for helping us to render several figures shown
in this paper. We acknowledge the support from the Supercomputing
Center of USTC. This work is supported by NSFC 10533050, NSFC
10703006 and NSFC 10973013.

\end{document}